\def\comp{{\rm C}\llap{\vrule height7.1pt width1pt depth-.4pt\phantom t}}
\def\Box{\kern1pt\vbox{\hrule height 1.2pt\hbox{\vrule width 1.2pt\hskip 3pt
   \vbox{\vskip 6pt}\hskip 3pt\vrule width 0.6pt}\hrule height 0.6pt}\kern1pt}
\def\gtwid{\mathrel{\raise.3ex\hbox{$>$\kern-.75em\lower1ex\hbox{$\sim$}}}}
\def\ltwid{\mathrel{\raise.3ex\hbox{$<$\kern-.75em\lower1ex\hbox{$\sim$}}}}
 
\documentstyle[epsfig,12pt]{article}

\newcommand{\be}{\begin{equation}}
\newcommand{\ee}{\end{equation}}

\newcommand{\vp}{\varphi}

\newcommand{\kp}{\kappa}

\begin{document}
\begin{titlepage}
\begin{flushright}
astro-ph/0109272 \\ UFIFT-HEP-00-22
\end{flushright}
\vspace{.4cm}
\begin{center}
\textbf{No One Loop Back-Reaction In Chaotic Inflation}
\end{center}
\begin{center}
L. R. Abramo$^{\dagger}$
\end{center}
\begin{center}
\textit{Theoretische Physik \\ Ludwig Maximilians Universit\"{a}t \\
Theresienstr. 37, \\ D-80333 M\"{u}nchen GERMANY}
\end{center}
\begin{center}
R. P. Woodard$^{\ddagger}$
\end{center}
\begin{center}
\textit{Department of Physics \\ University of Florida \\ 
Gainesville, FL 32611 USA}
\end{center}

\begin{center}
ABSTRACT
\end{center}

\hspace*{.5cm} We use an invariant operator to study the quantum gravitational 
back-reaction to scalar perturbations during chaotic inflation. Our
operator is the inverse covariant d'Alembertian expressed as a function 
of the local value of the inflaton. In the slow roll approximation this 
observable gives $-1/(2 H^2)$ for an arbitrary homogeneous and isotropic 
geometry, hence it is a good candidate for measuring the local expansion 
rate even when the spacetime is not perfectly homogeneous and isotropic. 
Corrections quadratic in the scalar creation and annihilation operators of
the initial value surface are included using the slow-roll and long 
wavelength approximations. The result is that all terms which could 
produce a significant secular back-reaction cancel from the operator,
before one even takes its expectation value. Although it is not relevant 
to the current study, we also develop a formalism for using stochastic 
samples to study back-reaction.
\begin{flushleft}
\end{flushleft}
\vspace{.4cm}
\begin{flushleft}
$^{\dagger}$ e-mail: abramo@theorie.physik.uni-muenchen.de \\
$^{\ddagger}$ e-mail: woodard@phys.ufl.edu
\end{flushleft}
\end{titlepage}

\section{Introduction}

Cosmological perturbations generated during an inflationary
era are almost certainly responsible for the density ripples 
out of which the presently observed stars, galaxies and clusters have 
grown. The basic inflationary mechanism that generated the seeds for 
galaxy formation and growth of large-scale structure is quantum 
mechanical particle production powered by the accelerated expansion 
of the Universe. Now that the latest balloon experiments \cite{CMBR} 
have all but validated this picture, we are left with the duty of 
testing further the consequences of particle production in the 
early Universe. An outstanding question is whether quantum pair 
creation has any effect on the background in which it takes place.

The simplest system in which the problem of back-reaction
can be studied is scalar-driven inflation.
Since dynamical scalars mix with the gravitational 
potentials they cause, gravitational back-reaction can in 
principle occur at one loop in scalar-driven inflation.
Indeed, previous results by the present authors and 
collaborators \cite{MAB,ABM,AbWo1,AbWo2} suggested that the 
one loop back-reaction of scalar perturbations could slow
down the expansion rate of the Universe in the case 
of the so-called ``chaotic'' inflationary models.

In \cite{MAB,ABM}, spatial averaging and a fixed gauge were employed 
in order to compute the effective energy-momentum tensor for cosmological 
perturbations. The present authors studied the same physical problem 
\cite{AbWo1,AbWo2}, this time taking expectation values of the metric 
and using those expectation values to form invariants, and came to 
conclusions identical to those of \cite{MAB,ABM}. Nevertheless, these 
works have been criticized in two grounds.

The first objection \cite{Unruh} concerns the use of expectation 
values of the gauge-fixed metric to build physical observables. The
concern is that certain variations in the gauge fixing condition 
could change the expectation value of the metric in ways 
which cannot be subsumed into a coordinate transformation. 
Therefore, forming the expectation value of the metric into 
coordinate invariant quantities would not purge these quantities 
of gauge dependence. Back-reaction should then be studied with 
an operator which is itself an invariant, before taking the 
expectation value.

The second objection \cite{Pritzker} states that using expectation 
values invites a Schr\"odinger Cat paradox. This is because 
the process of superadiabatic amplification leaves superhorizon modes in
highly squeezed states whose behavior is essentially classical. That is, the
various portions of the wave function no longer interfere with one another. 
Quantum mechanics determines the random choice of where we are in the wave
function, but the evolution of each portion is approximately classical. Under
these conditions, averaging over the full wave function does not give a good
representation of the physics. In the context of the Schr\"odinger Cat paradox
Linde observes: ``When you open the box a week later what you find is either a
very hungry cat or else a smelly piece of meat, not the average of the two.'' 
He and his colleagues would prefer that back-reaction be studied 
stochastically \cite{stoch}, where each mode is assigned a random, 
$\comp$-number value as it experiences horizon crossing and evolves 
classically thereafter. 

In the present study of the one-loop back-reaction effect we
have tried to address these objections. To avoid potential problems 
from using the gauge fixed metric we have instead computed the 
functional inverse of the covariant d`Alembertian (see also
\cite{Obs1} for a more detailed discussion):
\begin{equation}
\label{CD}
\Box_c \equiv 
\frac{1}{\sqrt{-g}} \partial_{\mu} \left(\sqrt{-g} g^{\mu\nu} 
\partial_{\nu} \right) - \frac16 R \; .
\end{equation}
This operator --- ${\cal{A}} \equiv \Box_c^{-1}[g](t,{\vec x})$ --- 
averages over the past lightcone, very much as astronomers do when 
compiling a Hubble diagram. In the slow roll approximation it gives 
$-\frac12 H^{-2}$ for an arbitrary homogeneous and isotropic universe 
\cite{Obs1}. The scalar ${\cal{A}}$ is therefore a reasonable candidate 
for measuring the local expansion rate when the universe is not precisely 
homogeneous and isotropic. By measuring ${\cal A}$ on 3-surfaces upon
which the dynamical scalar is constant we promote it into a full invariant,
${\cal A}_{\rm inv}$.

At one loop, and to leading order in the standard infrared expansion, 
corrections to the operator ${\cal A}$ depend upon a variable we call
$\Phi(x)$. This is the accumulated Newtonian potential from infrared 
modes. $\Phi(x)$ grows with time, since more and more modes redshift 
into the infrared regime as inflation progresses, so dependence upon 
any positive power of $\Phi(x)$ would give a secular effect. However, 
when full invariance is enforced by evaluating ${\cal A}$ on 3-surfaces 
of constant scalar, all terms involving only the undifferentiated 
Newtonian potential cancel. This result was anticipated by Unruh \cite{Unruh}
who noted that the linearized mode functions become pure gauge in the long
wavelength limit. The fact that more and more modes redshift to the infrared
regime is not pure gauge, but Unruh's observation means that physical effects 
must involve derivatives of $\Phi(x)$. Such terms do contribute, but they are
small.  We conclude that there is no significant deviation from the background 
expansion rate at this order, in marked contrast with the results of 
\cite{MAB,ABM,AbWo1}.

Section 2 describes the perturbative background in which we work. It also
illustrates the slow roll expansion through which one can perform most 
operations of temporal calculus in an arbitrary inflationary background. 
Section 3 describes the formalism by which the field equations are solved 
perturbatively. It concludes with the long wavelength approximation which 
effectively removes spatial dependence and produces a one dimensional
problem that can be treated in the slow roll expansion. Section 4 applies 
the technology to determine the various field operators to quadratic order
in the creation and annihilation operators on the initial value surface.
The operator ${\cal A}$ is evaluated to the same order in Section 5.
The distinction between expectation values and stochastic samples is 
irrelevant, at this order, in view of the cancellation of all potentially
significant one loop corrections to the ${\cal A}_{\rm inv}$ operator. However,
the issue should re-emerge at higher loops where secular back-reaction 
can derive from the coherent superposition of interactions within the 
observer's past lightcone. We have therefore developed a formalism for the
perturbative study of stochastic effects, which is presented in Section 6.
Our results are summarized and discussed in Section 7.

\section{The perturbative background}

The system under study is that of general relativity with a general,
minimally coupled scalar:
\begin{equation}
{\cal L} = {1 \over 16\pi G} R \sqrt {-g} - \frac12 \partial_{\mu} 
\varphi \partial_{\nu} \varphi g^{\mu \nu} \sqrt {-g} - V(\varphi) 
\sqrt {-g} \; . 
\label{eq:action}
\end{equation}
This section concerns the homogeneous and isotropic backgrounds $g_0$ and 
$\varphi_0$ about which perturbation theory will be formulated. Three 
classes of identities turn out to be interesting for our purposes:

\begin{enumerate}
\item Those which are exact and valid for any potential $V(\varphi)$;
\item Those which are valid in the slow roll approximation but still for 
any potential; and 
\item Those which are valid for the slow roll approximation with the 
potential $V(\varphi) = \frac12 m^2 \varphi^2$.
\end{enumerate}
We shall develop them in this order, identifying the point at which each
further specialization and approximation is made.

Among the exact identities is the relation between co-moving and conformal
coordinates:
\begin{equation}
ds_0^2 = -dt^2 + a^2 d{\vec x} \cdot d{\vec x} = a^2(\eta) 
\left\{-d\eta^2 + d{\vec x} \cdot d{\vec x}\right\} \; ,
\end{equation}
where conformal time $\eta$ is defined in the usual way:
\begin{equation}
dt = a (\eta) \, d\eta \quad \; .
\end{equation}
The Hubble ``constant'' is the logarithmic co-moving time derivative of the 
background scale factor:
\begin{equation}
H \equiv {\dot{a} \over a} = {a^{\prime} \over a^2} \; ,
\end{equation}
where a dot denotes differentiation with respect to (background) co-moving
time and a prime stands for differentiation with respect to conformal time.

Two of Einstein's equations are nontrivial in this background:
\begin{eqnarray}
\label{G00}
3 H^2 & = & \frac12 \kappa^2 \left\{\frac12 \dot{\varphi}_0^2 + V(\varphi_0)
\right\} \; ,
\\ \label{Gij}
-2 \dot{H} - 3 H^2 & = & \frac12 \kappa^2 \left\{\frac12 \dot{\varphi}_0^2
- V(\varphi_0)\right\} \; ,
\end{eqnarray}
where $\kappa^2 \equiv 16 \pi G$ is the loop counting parameter of perturbative
quantum gravity. By adding the two Einstein equation one can solve for the time
derivative of the scalar, which we assume to be negative:
\begin{equation}
\label{eq:fidot}
\dot{\varphi}_0 = -{2 \over \kappa} \sqrt{-\dot{H}} \; .
\end{equation}

Successful models of inflation require the following two conditions which 
define the {\it slow roll approximation}:
\begin{eqnarray}
\vert \ddot{\varphi}_0 \vert & \ll & H \vert \dot{\varphi}_0 \vert \; ,
\label{eq:slow1} \\
\dot{\varphi}_0^2 & \ll & V(\varphi_0) \; . 
\label{eq:slow2}
\end{eqnarray}
It follows that there are two small parameters. Although these are 
traditionally expressed as ratios of the potential and its derivatives the more
useful quantities for our work are ratios of the Hubble constant and its 
derivatives:
\begin{equation}
{-\dot{H} \over H^2} \ll 1 \qquad , \qquad {\vert \ddot{H} \vert \over - H 
\dot{H}} \ll 1 \; .
\end{equation}
For models of interest to us the rightmost of these parameters is negligible 
with respect to the leftmost one.

The slow roll approximation gives useful expansions for simple calculus
operations. For example, ratios of derivatives of the field are:
\begin{eqnarray}
\label{eq:idfi1}
{\varphi_0^{\prime\prime}\over \varphi_0^{\prime}} & = & a H \left( 1 +
\frac12 {\ddot{H} \over H \dot{H}} \right) \; ,\\
\label{eq:idfi2}
{\varphi_0^{\prime\prime\prime}\over \varphi_0^{\prime}} & = & 2 a^2 H^2 
\left(1 + {\dot{H} \over 2 H^2} + \dots\right) \; .
\end{eqnarray}
Successive partial integration also defines useful slow roll expansions:
\begin{eqnarray}
\label{int1}
\int dt H^{\alpha} a^{\beta} & = & {1 \over \beta} H^{\alpha-1} 
a^{\beta} \left\{1 + {(\alpha - 1) \over \beta} \left({-\dot{H} \over H^2}
\right) + \dots\right\} \; ,\\ \label{int2}
\int dt H^{\alpha} & = & {1 \over \alpha + 1} {H^{\alpha+1} \over \dot{H}}
\left\{1 + {1 \over \alpha + 2} {H \ddot{H} \over \dot{H}^2}+ \dots\right\} 
\; .
\end{eqnarray}

\section{Einstein-scalar during inflation}

The purpose of this section is to describe our procedure for expressing the
metric and scalar fields at any spacetime point as functionals of the
unconstrained initial value operators. These unconstrained initial value
operators are the only true degrees of freedom in any quantum theory and
expressing the dynamical variables as functionals of them is what it means
to solve the equations of motion in the Heisenberg picture. As a simple
example, if the dynamical variable is the position $q(t)$ of a one-dimensional
particle then the equation of motion,
\begin{equation}
\ddot{q}(t) + \omega^2 q(t) = 0 \; ,
\end{equation}
has as its ``solution'',
\begin{equation}
q(t) = q_0 \cos(\omega t) + {\dot{q}_0 \over \omega} \sin(\omega t) \; .
\end{equation}
In this case the unconstrained initial value operators are $q_0$ and $\dot{q}_0$.

Obtaining such a solution perturbatively entails two expansions. In the first 
stage one writes the dynamical variables as background plus perturbations
and expands the equations of motion in powers of the perturbed quantities.
To continue with our particle example, suppose the full equation of motion is
$E[x](t) = 0$ and that we are perturbing around some classical solution $X(t)$.
If we define the perturbed position as $q(t)$ then $x(t) \equiv X(t) + q(t)$. 
Since the background is a solution the perturbative equations of motion 
necessarily begin at linear order,
\begin{equation}
E[X + q](t) = \ddot{q}(t) + \omega^2 q(t) + {\Delta E}[q](t) \; ,
\end{equation}
where ${\Delta E}[q](t)$ contains terms of order $q^2$ and higher.

The second stage of perturbation theory consists of solving for the perturbed
quantities as a series in powers of the unconstrained initial value operators.
This is done by first solving the linearized equations so as to make the 
unconstrained variables agree with their full initial value data on the initial
value surface. The linearized solution in our particle example would be,
\begin{equation}
q_{\rm lin}(t) = \sqrt{\hbar \over 2 m \omega} \left\{a e^{-i \omega t} +
a^{\dagger} e^{i \omega t}\right\} \; ,
\end{equation}
where we have chosen to organize the initial value data into the annihilation
operator, $a \equiv \sqrt{m \omega/2\hbar} (q_0 + i\dot{q}_0/\omega)$, and its
conjugate.

One then ``integrates'' the full equations of motion using the retarded Green's
functions of the linearized theory. The retarded Green's function of our 
particle example is,
\begin{equation}
G_{\rm ret}(t;t') = {\theta(t-t') \over \omega} \sin[\omega (t - t')] \; ,
\end{equation}
and the integrated equation of motion is,
\begin{equation}
q(t) = q_{\rm lin}(t) - \frac1{\omega} \int_0^t dt' \sin[\omega (t-t')] 
{\Delta E}[q](t') \; .
\end{equation}
The perturbative solution derives from successive substitution of this
integrated form for the perturbed quantities on the righthand side of the
equation,
\begin{equation}
q(t) = q_{\rm lin}(t) -\frac1{\omega} \int_0^t dt' \sin[\omega (t-t')]
{\Delta E}\left[q_{\rm lin}\right](t') + \dots \; .
\end{equation}

It will be seen that the perturbative operator expansion which results bears
a close relation to the diagrammatic expansion of quantum field theoretic
amplitudes. The place of external lines is taken by the linearized solution;
propagators are replaced by the corresponding retarded Green's functions; and,
except for some factors of $i$, there is no change at all in the vertices.
Since no loop diagrams appear there is no need of Faddeev-Popov ghosts. 
Owing to this correspondence many of the results we shall obtain for the 
operator expansion have already appeared in our previous perturbative 
expansion for expectation values \cite{AbWo1}. We shall retain most of the 
conventions used in that paper.

In a gauge theory such as gravity there is the additional complication of
constraints which require some of the dynamical variables to be nonlinear
functionals of the unconstrained initial value operators even on the initial
value surface. We shall deal with this by working in a gauge with a nonlocal
field redefinition such that there are no nonlinear corrections on the 
initial value surface. Of course the nonlinear corrections required by the
initial value constraints return when our field redefinition is inverted to
give the perturbed metric and scalar fields.

Another complication is that one cannot generally obtain explicit solutions 
to the linearized field equations in the time dependent backgrounds of 
chaotic inflation. We therefore explain how to construct plane wave solutions 
as series expansions around the ultraviolet and infrared limiting cases. 
Since the putative physical effect derives from modes which have redshifted 
into the infrared regime it is the long wavelength approximation --- coupled 
with the slow roll expansion --- which allows us to obtain quantitative results.

Finally, it should be noted that we are not actually including all the
physical degrees of freedom. Since gravitons cannot give a significant
back-reaction at one loop we have suppressed the initial value operators
associated with them, leaving only scalar degrees of freedom. Of course 
we still work out how the full metric depends upon this scalar initial 
value data.

\subsection{Perturbative field equations}

The field equations derived from the Einstein-scalar action (\ref{eq:action}) 
are most usefully expressed in the form,
\begin{eqnarray}
F & \equiv & {1 \over a} \left( \Box \vp - V'(\vp) \right) = 0 \; , \\
E^{\mu\nu} & \equiv & -{a \over \kappa} \left(G^{\mu\nu} - \frac{\kappa^2}2
T^{\mu\nu}\right) = 0 \; . \label{eq:Einstein}
\end{eqnarray}
Here $G^{\mu\nu} \equiv R^{\mu\nu} - \frac12 g^{\mu\nu} R$ is the Einstein
tensor and $T^{\mu\nu}$ is the scalar stress tensor whose covariant expression
has the form,
\begin{equation}
T_{\mu\nu} \equiv \partial_{\mu} \vp \partial_{\nu} \vp - g_{\mu\nu} \left(
\frac12 g^{\rho\sigma} \partial_{\rho} \vp \partial_{\sigma} \vp + V(\vp)\right)
\; .
\end{equation}
Our dynamical variables are the scalar perturbation $\phi$ and the 
conformally rescaled pseudo-graviton $\psi_{\mu\nu}$:
\begin{eqnarray}
\label{def:scalar}
\vp(\eta,\vec{x}) & \equiv & \vp_0(\eta) + \phi(\eta,\vec{x})
\\ \label{def:metric}
g_{\mu\nu}(\eta,\vec{x}) & \equiv & a^2(\eta) \left[ \eta_{\mu\nu} + \kp 
\psi_{\mu\nu}(\eta,\vec{x}) \right] \; ,
\end{eqnarray}
where $\eta_{\mu\nu}$ is the (spacelike) Lorentz metric which is used to
raise and lower pseudo-graviton indices. We define the quantities ${\Delta F}$ 
and ${\Delta E}^{\mu \nu}$ to include all terms of second and higher orders in 
$\psi_{\mu\nu}$ and $\phi$. One can read off the quadratic terms of ${\Delta F}$ 
and ${\Delta E}^{\mu\nu}$ from Tables 1 and 3 of \cite{AbWo1}: multiply the
terms of Table 1 by $1/a$ and then drop the ``external'' $\psi_{\mu\nu}$; 
vary the terms of Table 3 with respect to $\phi$ or $\psi_{\mu\nu}$ and then
multiply by $1/a$.

The linearized equations are vastly simplified by imposing the gauge
condition,
\begin{equation}
\label{gaugefix}
F_{\mu} \equiv a \left(\psi_{\mu~ ,\nu}^{~\nu} - \frac12 \psi_{,\mu} - 2 
{a^{\prime} \over a} \psi_{\mu 0} + \eta_{\mu 0} \kappa 
{\varphi}_0^{\prime} \phi \right) = 0 \; ,
\end{equation}
where a comma denotes ordinary differentiation and $\psi \equiv \eta^{\mu\nu}
\psi_{\mu\nu}$. It is also useful to $3+1$ decompose and to slightly rearrange 
the perturbed fields as follows,
\begin{equation}
f \equiv  a \phi \quad , \quad z \equiv a \psi_{00} \quad , \quad v_i \equiv 
a \psi_{0i} \quad , \quad h_{ij} \equiv a \left( \psi_{ij} - \delta_{ij} 
\psi_{00} \right) \; . \label{eq:vars}
\end{equation}
The various differential operators of the linearized equations can also
be given a simple common form,
\begin{equation}
\widehat{\cal{D}}_I \equiv \nabla^2 + {\cal{D}}_I \equiv \nabla^2 
- \partial_0^2 + \frac{\theta_I''(\eta)}{\theta_I(\eta)} \; , \label{def:DI}
\end{equation}
where $I=A,B,C$ and
\be
\label{thetaI}
\theta_A \equiv a \quad , \quad \theta_B \equiv a^{-1} \quad , \quad 
\theta_C \equiv \frac{1}{a} \sqrt{H^2 \over -\dot{H}} \; .
\ee

Making use of the gauge condition results in the following expansions
for $F$ and the various components of $E^{\mu\nu}$ \cite{AbWo1},
\begin{eqnarray}
F & = & -\kappa \varphi_0^{\prime\prime} z + \left(\widehat{\cal D}_B + 
{\varphi_0^{\prime \prime\prime} \over \varphi_0^{\prime}}\right) f 
+ {\Delta F} \; , \label{eq:second} \\
E^{00} & = & \widehat{\cal D}_B z - \kappa \varphi_0^{\prime\prime} f + 
\frac14 D_A h + {\Delta E}^{00} \; , \label{eq:first} \\
E^{0i} & = & -\frac12 \widehat{\cal D}_B v_i + {\Delta E}^{0i} \; , 
\label{eq:half} \\ 
E^{ij} & = & \frac12 \widehat{\cal D}_A \left(h_{ij} - \frac12 \delta_{ij} 
h\right) + {\Delta E}^{ij} \; . \label{eq:zero} 
\end{eqnarray}
The mixing in (\ref{eq:zero}) between $h_{ij}$ and its trace $h \equiv h_{kk}$ 
can be removed by taking linear combinations. The same is true for the 
mixing in (\ref{eq:first}) between $h$ and the variables $f$ and $z$. 
However, the mixing between $z$ and $f$ in (\ref{eq:second}-\ref{eq:first}) 
cannot be removed algebraically for a general background. To diagonalize this
sector we must make a nonlocal field redefinition,
\begin{eqnarray}
x(\eta,{\vec x}) & \equiv & z^{\prime}(\eta,{\vec x}) + \frac{\kappa}2 
\varphi_0^{\prime}(\eta) f(\eta,{\vec x}) \; , \label{eq:xdef} \\
y(\eta,{\vec x}) & \equiv & \frac{\kappa}2 \varphi_0^{\prime}(\eta) z(\eta,
{\vec x}) + \varphi_0^{\prime}(\eta) \left({f(\eta,\vec{x}) \over 
\varphi_0^{\prime}(\eta)}\right)^{\prime} \; . \label{eq:ydef}
\end{eqnarray}
When all this is done the fully diagonalized equations take the form,
\begin{eqnarray}
\widehat{\cal D}_A x & = & -\left({\Delta E}^{00} + {\Delta E}^{ii}
\right)^{\prime} - \frac{\kappa}2 \varphi_0^{\prime} {\Delta F} \; ,
\label{eq:x} \\
\widehat{\cal D}_C y & = & -\frac{\kappa}2 \vp_0^{\prime} \left(
{\Delta E}^{00} + {\Delta E}^{ii} \right) - \varphi_0^{\prime} 
\left({{\Delta F} \over \varphi_0^{\prime}} \right)^{\prime} \; , 
\label{eq:y} \\
\widehat{\cal D}_B v_i & = & 2 {\Delta E}^{0i} \; , \label{eq:v} \\ 
\widehat{\cal D}_A h_{ij} & = & -2 \left({\Delta E}^{ij} -\delta_{ij}
{\Delta E}^{kk}\right) \; . \label{eq:h} 
\end{eqnarray}

Finally, one must check that the gauge conditions can be consistently
imposed as constraints. Their $3 + 1$ decomposition in the diagonal 
variables takes the form,
\begin{equation}
F_0 = -2 x + v_{i,i} - \frac{a}2 \left({h \over a}\right)^{\prime}
\quad , \quad F_i = -\frac1{a} \left(a v_i\right)^{\prime} - h_{ij,j} 
- \frac12 h_{,i} \; . 
\end{equation}
Using the ``commutator'' identities,
\begin{equation}
\widehat{\cal D}_B a \partial_0 \frac1{a} = a \partial_0 \frac1{a}
\widehat{\cal D}_A \quad , \quad \widehat{\cal D}_A \frac1{a} \partial_0 a 
= \frac1{a} \partial_0 a \widehat{\cal D}_B \; ,
\end{equation}
and the gauge-fixed field equations (\ref{eq:x}-\ref{eq:h}), it is easy
to show that the constraints obey,
\begin{eqnarray}
\widehat{\cal D}_B F_0 & = & 2 \left({\Delta E}^{\mu 0}_{~~ , \mu} + 
\frac{a'}{a} {\Delta E}^{ii} + \frac{\kappa}2 \varphi^{\prime}_0 
{\Delta F}\right) \; , \label{eq:bianchi1} \\
\widehat{\cal D}_A F_i & = & -2 \left({\Delta E}^{\mu i}_{~~ , \mu} + 
\frac{a'}{a} {\Delta E}^{0 i}\right) \; . \label{eq:bianchi2}
\end{eqnarray}
The quantities on the right hand side of (\ref{eq:bianchi1}-\ref{eq:bianchi2})
vanish as a consequence of the background Bianchi identities in the usual
way. Therefore, the constraints are preserved by the gauge fixed evolution
equations and full equivalence between the invariant field equations and those
of the gauge-fixed theory will hold if the initial value operators are constrained
to make $F_{\mu}$ and its first conformal time derivative vanish at $\eta = \eta_0$.

\subsection{Unconstrained plane waves}

It is useful to consider plane wave solutions to the linearized, gauge-fixed 
field equations before suppressing the physical gravitons and using the 
constraints and the residual gauge freedom to purge unphysical initial value 
data. That is, we seek the kernel of the differential operators ${\cal D}_I - k^2 
= -\partial_0^2 - k^2 + \theta_I^{\prime\prime}/\theta_I$, where the functions 
$\theta_I(\eta)$ were defined in relation (\ref{thetaI}). Although explicit 
solutions cannot be obtained for an arbitrary inflationary background, the 
limiting cases of $k = 0$ and $\theta_I = 0$ serve as the basis of series 
solutions for the ``infrared'' ($k^2 \ll \theta_I^{\prime \prime}/\theta_I$) 
and ``ultraviolet'' ($k^2 \gg \theta_I^{\prime\prime}/\theta_I$) regimes. 
In the ultraviolet regime the particle interpretation is the same as for 
flat space so its normalization is used to define that of the infrared 
regime. The final step of this subsection is superposing plane waves 
multiplied by the creation and annihilation operators of the linearized, 
gauge-fixed action. The commutator functions between such fields give 
the retarded Green's functions needed to integrate the perturbative 
field equations.

The following two solutions comprise a useful basis set in the infrared limit 
($k = 0$),\footnote{Here our notation differs slightly from \cite{AbWo1} in 
that the signs of $\theta_C$ and of the second solutions Eq. (\ref{modsol2}) 
have been inverted.}:
\begin{equation}
Q_{10,I}(\eta) = \theta_I(\eta) \quad , \quad Q_{20,I}(\eta) = 
\theta_I(\eta) \int^\eta_{-\infty} d\eta' \frac{1}{\theta_I^2(\eta')} \; . 
\label{modsol2}
\end{equation}
One of these is a ``growing'' mode, the other a ``decaying'' mode. For example, 
the decaying $C$ mode is $Q_{10,C}(\eta)$. The growing $C$ mode can be 
evaluated in the slow roll expansion,
\be
\label{growing}
Q_{20,C}(\eta) = \frac{\sqrt{-\dot{H}}}{H^2} \left[ 1 + {\cal{O}} 
\left( \frac{-\dot{H}}{H^2} \right) \right] \; .
\ee

We can find the inverse of ${\cal{D}}_I$ on any function $f(\eta)$ by two
simple integrations,
\be
\label{invD}
\left({\cal D}_I^{-1} f \right) (\eta) = - \theta_I (\eta) 
\int_{\eta_0}^\eta d\eta' \frac{1}{\theta^2_I(\eta')} \int_{\eta_0}^\eta 
d\eta'' \theta_I (\eta'') f(\eta'') \; . 
\ee
By integrating the full equation, ${\cal D}_I Q = k^2 Q$, one obtains a
relation,
\be
\label{irsol}
Q_{i,I} (\eta,k) = Q_{i0,I}(\eta) + k^2 \left({\cal D}_I^{-1} Q_{i,I}
\right)(\eta,k) \quad i=1,2 \; ,
\ee
whose iteration results in a convergent series expansion in powers of $k^2$.
Note that constancy of the Wronskian and the vanishing of corrections at 
$\eta = \eta_0$ allows us to evaluate it using the zeroth order solutions,
\be
\label{wronskianIR}
Q_{2,I}'(\eta,k) Q_{1,I} (\eta,k) - Q_{1,I}'(\eta,k) Q_{2,I} (\eta,k)= 1 \; .
\ee

The useful basis elements for the ultraviolet limit ($\theta_I = 0$) are 
$e^{-i k \eta}$ and its conjugate. In this limit we know the absolute 
normalization from correspondence with flat space. Coupling this with the 
harmonic oscillator Green's function gives the integrated mode equation,
\be
Q_I(\eta,k) = \frac{e^{-ik\eta}}{\sqrt{2k}} + \frac1{k} \int^\eta_{-\infty} 
d\bar{\eta} \, \sin{[k(\eta-\bar{\eta})]} \frac{\theta_I''(\bar{\eta})}{
\theta_I(\bar{\eta})} Q_I (\bar{\eta},k) \; ,
\label{UV}
\ee
whose iteration yields an asymptotic expansion in powers of $1/k$. The 
Wronskian for the ultraviolet mode functions is,
\be
Q_{I}'(\eta,k) Q_{I}^* (\eta,k) - {Q_{I}'}^*(\eta,k) Q_{I} (\eta,k) = -i \; .
\label{wronskianUV}
\ee

Any quantum operator $\psi_I(\eta,\vec{x})$ which is annihilated by 
$\widehat{\cal D}_I$ can be expressed as a superposition of plane waves 
with operator coefficients,
\begin{equation}
\psi_I(\eta,\vec{x}) = \int {d^3k \over (2\pi)^3} \left\{ Q_I(\eta,k) 
e^{i \vec{k} \cdot \vec{x}} \Psi(\vec{k}) + Q_I^*(\eta,k) e^{-i \vec{k} 
\cdot \vec{x}} \Psi^{\dagger}(\vec{k}) \right\} \; .
\end{equation}
If the conjugate momentum is $\psi_I^{\prime}(\eta,\vec{x})$ then the 
Wronskian (\ref{wronskianUV}) implies that the coefficients have the 
algebra of canonically normalized creation and annihilation operators,
\begin{equation}
\left[\Psi(\vec{k}), \Psi^{\dagger}(\vec{k}^{\prime})\right] = 
\delta{ij} (2\pi)^3 \delta^3(\vec{k} - \vec{k}^{\prime}) \; .
\end{equation}
By taking the commutator of two such fields we obtain a sequence of lovely 
expressions for the retarded Green's function of the differential operator 
$\widehat{\cal D}_I$,
\begin{eqnarray}
\lefteqn{G_I(x;x') = -i \theta({\Delta \eta}) \left[\psi_I(\eta,\vec{x}),
\psi_I(\eta',\vec{x}')\right] \; ,} \\
& & = -i \theta({\Delta \eta}) \int {d^3k \over (2\pi)^3} e^{i \vec{k} \cdot
{\Delta \vec{x}}} \left\{ Q_I(\eta,k) Q_I^*(\eta',k) - (\eta \leftrightarrow
\eta')\right\} \; , \label{GrI} \\
& & = \theta({\Delta \eta}) \int {d^3k \over (2\pi)^3} e^{i \vec{k} \cdot
{\Delta \vec{x}}} \left\{ Q_{1,I}(\eta,k) Q_{2,I}(\eta',k) - (\eta 
\leftrightarrow \eta')\right\} \; ,
\end{eqnarray}
where ${\Delta \eta} \equiv \eta - \eta'$ and ${\Delta \vec{x}} \equiv
\vec{x} - \vec{x}'$. The crucial final expression --- in terms of the infrared 
mode functions --- follows from the fact that that it obeys the same differential 
equation --- $\widehat{\cal D}_I G_I(x;x') = \delta^4(x-x')$ --- and retarded 
boundary conditions as the first two.

The preceding analysis suffices to define the retarded Green's functions
needed to integrate equations (\ref{eq:v}) and (\ref{eq:h}),
\begin{eqnarray}
v_i(\eta,\vec{x}) & = & v^{\rm lin}_i(\eta,\vec{x}) + 2 \int_{\eta_0}^{\eta}
d\eta' \int d^3x' G_B(x;x') {\Delta E}^{0i}(x') \; , \label{eq:vint} \\ 
h_{ij}(\eta,\vec{x}) & = & h^{\rm lin}_{ij}(\eta,\vec{x}) \nonumber \\
& & \; -2 \left(\delta_{ik} \delta_{j\ell} - \delta_{ij} \delta_{k\ell}\right) 
\int_{\eta_0}^{\eta} d\eta' \int d^3x' G_A(x;x') {\Delta E}^{k\ell}(x') \; . 
\qquad \label{eq:hint} 
\end{eqnarray}
It would be straightforward to give plane wave expansions for $v^{\rm lin}_i$
and $h^{\rm lin}_{ij}$. However, we will not bother since these linearized
fields vanish when physical graviton degrees of freedom are suppressed and the 
various constraints and residual gauge conditions are imposed. 

The same general technology can be applied to integrate the more complicated
$(f,z)$ system,
\begin{equation}
\left( \begin{array}{cc} \hat{\cal{D}}_B + \frac{\vp_0'''}{\vp_0'} & -\kp\vp_0'' \\
-\kp\vp_0'' & \hat{\cal{D}}_B \end{array} \right)
\left( \begin{array}{c} f \\ z \end{array} \right) = -
\left( \begin{array}{c} {\Delta F} \\ {\Delta E}^{00} + {\Delta E}^{ii} 
\end{array} \right) \; . \label{eq:zfsys}
\end{equation}
The desired retarded Green's functions,
\begin{equation}
\left( \begin{array}{cc} \hat{\cal{D}}_B + \frac{\vp_0'''}{\vp_0'} & -\kp\vp_0'' \\
-\kp\vp_0'' & \hat{\cal{D}}_B \end{array} \right) 
\left( \begin{array}{cc} G_{ff}(x;x') &  G_{fz}(x;x') \\  
G_{zf}(x;x') &  G_{zz}(x;x') \end{array} \right) 
= \delta^4(x-x') \left( \begin{array}{cc} 1 &  0 \\  
0 & 1 \end{array} \right) \; , \label{eq:gfsys}
\end{equation}
also follow from commutators of the linearized fields,
\begin{equation}
\left( \begin{array}{cc} G_{ff}(x;x') &  G_{fz}(x;x') \\  
G_{zf}(x;x') &  G_{zz}(x;x') \end{array} \right) = - i \theta(\Delta \eta)
\left( \begin{array}{cc} \left[ f(x),f(x') \right] & \left[ f(x),z(x') \right] \\
\left[ z(x),f(x') \right] & \left[ z(x),z(x') \right] \end{array} \right)
\; . \label{eq:ggfs}
\end{equation}
Unfortunately it is the fields $x(\eta,\vec{x})$ and $y(\eta,\vec{x})$ which
are annihilated by $\widehat{\cal D}_B$ and $\widehat{\cal D}_C$, so it is
they that possess simple plane wave expansions in terms of the $B$ and $C$
mode functions.

At linearized order the gauge-fixed action reveals the conjugate momenta to
$f$ and $z$ to be simply their conformal time derivatives. Definitions 
(\ref{eq:xdef}) and (\ref{eq:ydef}) give the following nonzero equal-time
commutation algebra for $x$ and $y$,
\begin{equation}
\left[x(\eta,\vec{x}), x'(\eta,\vec{y})\right] = -i \nabla^2 \delta^3(\vec{x} 
- \vec{y}) = \left[y(\eta,\vec{x}), y'(\eta,\vec{y})\right] \; . \label{xycom}
\end{equation}
If we still employ canonically normalized creation and annihilation 
operators,
\begin{equation}
\left[X(\vec{k}), X(\vec{k}')\right] =  (2\pi)^3 \delta^3(\vec{k} - \vec{k}') 
= \left[Y(\vec{k}), Y(\vec{k}')\right] \; ,
\end{equation}
then the additional factor of $-\nabla^2$ in (\ref{xycom}) implies the 
following plane wave expansions,
\begin{eqnarray}
x(\eta,{\vec x}) & = & \int {d^3k \over (2\pi)^3} k \left\{ Q_B(\eta,k)
e^{i \vec{k} \cdot \vec{x}} X({\vec k}) + Q_B^*(\eta,k) e^{-i \vec{k} \cdot 
\vec{x}} X^{\dagger}({\vec k}) \right\} \; ,\\
y(\eta,{\vec x}) & = & \int {d^3k \over (2\pi)^3} k \left\{ Q_C(\eta,k)
e^{i \vec{k} \cdot \vec{x}} Y({\vec k}) + Q_C^*(\eta,k) e^{-i \vec{k} \cdot 
\vec{x}} Y^{\dagger}({\vec k}) \right\} \; .
\end{eqnarray}

Since the transformation between $(f,z)$ and $(x,y)$ involves conformal 
time derivatives, its inverse cannot be local in time for the off shell fields.
However, for the on-shell solutions we can use the linearized equations to
obtain the following expressions,
\begin{eqnarray}
f & = & {1 \over \nabla^2} \left[-\frac{\kappa}2 \varphi_0^{\prime} x + 
y^{\prime} + {\varphi_0^{\prime\prime} \over \varphi_0^{\prime}} y\right] \; .
\label{eq:ftrans} \\
z & = & {1 \over \nabla^2} \left[x^{\prime} - \frac{\kappa}2 
\varphi_0^{\prime} y \right] \; , \label{eq:ztrans} 
\end{eqnarray}
It is simplest to exploit these relations in evaluating the various commutators
of the $(f,z)$ retarded Greens functions. One then replaces products of the mode 
functions with the $B$ and $C$-type Green's functions. The answer is,
\begin{eqnarray}
\lefteqn{G_{ff}(x;x') = \frac{1}{\nabla^2} \left\{- \frac{\kp^2}4 \vp_0'(\eta) 
\vp_0'(\eta') G_B(x;x') \right.} \quad \nonumber \\
& & \qquad - \left. \frac1{\vp_0'(\eta) \vp_0'(\eta')} \partial_0 \partial_0'
\vp_0'(\eta) \vp_0'(\eta') G_C (x;x') + \delta^4(x-x') \right\} \; . 
\label{eq:ff} \\
\lefteqn{G_{fz}(x;x') = \frac{\kp}{2 \nabla^2} \left\{\vp_0'(\eta) 
\partial_0' G_B(x;x') + \frac{\vp_0'(\eta')}{ \vp_0'(\eta)} \partial_0 
\vp_0'(\eta) G_C (x;x') \right\} \; ,} \quad \label{eq:fz} \\ 
\lefteqn{G_{zf}(x;x') = \frac{\kp}{2 \nabla^2} \left\{\vp_0'(\eta') 
\partial_0 G_B(x;x') + \frac{\vp_0'(\eta)}{ \vp_0'(\eta')} \partial_0' 
\vp_0'(\eta') G_C (x;x') \right\} \; ,} \quad \label{eq:zf} \\
\lefteqn{G_{zz}(x;x') =} \quad \nonumber \\
& & \frac{1}{\nabla^2} \left\{-\partial_0 \partial_0' G_B(x;x') - \frac{\kp^2}4
\vp_0'(\eta) \vp_0'(\eta') G_C (x;x') + \delta^4(x-x') \right\} \; . \label{eq:zz} 
\end{eqnarray}
Finally, we substitute the relation, $G_I(x;x') = \widehat{\cal{D}}_I^{-1} 
\delta^4(x-x')$,
\begin{eqnarray}
G_{ff}(x;x') &=& \frac{1}{\nabla^2} \left[1 - \frac{\kp^2}{4} \vp_0' 
\widehat{\cal{D}}_B^{-1} \vp_0' + \frac{1}{\vp_0'} \partial_0 \vp_0' 
\widehat{\cal{D}}_C^{-1} \vp_0' \partial_0 \frac{1}{\vp_0'} \right] \delta^4(x-x') 
\; . \qquad \label{eq:ffD} \\
G_{fz}(x;x') &=& \frac{\kappa}{2 \nabla^2} \left[- \vp_0' \widehat{\cal{D}}_B^{-1} 
\partial_0 + \frac{1}{ \vp_0'}  \partial_0 \vp_0' \widehat{\cal{D}}_C^{-1} \vp_0'
\right] \delta^4 (x-x') \; , \label{eq:fzD} \\ 
G_{zf}(x;x') &=& \frac{\kappa}{2 \nabla^2} \left[ \partial_0 \widehat{\cal{D}}_B^{-1} 
\vp_0' - \vp_0' \widehat{\cal{D}}_C^{-1} \vp_0' \partial_0 \frac{1}{ \vp_0'} 
\right] \delta^4 (x-x') \; , \label{eq:zfD} \\ 
G_{zz}(x;x') &=& \frac{1}{\nabla^2} \left[ 1 + \partial_0 \widehat{\cal{D}}_B^{-1} 
\partial_0 - \frac{\kp^2}4 \vp_0' \widehat{\cal{D}}_C^{-1} \vp_0' \right] 
\delta^4(x-x') \; , \label{eq:zzD}
\end{eqnarray}

\subsection{Integrated field equations}

In this subsection we first purge unphysical initial value data by 
imposing the constraints and making use of the residual gauge freedom. 
Then gravitons are suppressed to leave what we call the physical 
linearized fields. Finally, the various field equations are integrated 
using the retarded Green's functions of the previous section.

Since our gauge condition involves derivatives it is preserved by residual
gauge transformations which obey second order differential equations.\footnote{
The analysis is quite similar to that carried out for de Sitter background in 
\cite{TsWo}.} One can therefore freely impose four residual conditions and 
their first conformal time derivatives on the initial value surface. We choose 
these to null $x(\eta_0,\vec{x})$, $x'(\eta_0,\vec{x})$, $v_i(\eta_0,\vec{x})$ 
and $v_i^{\prime}(\eta_0,\vec{x})$.

At the end of subsection 3.1 we showed that the four constraints are also
annihilated by second order differential operators. It therefore follows that
the initial values of the four constraints and their first conformal time
derivatives are the only extra conditions to be imposed in addition to the
gauge-fixed field equations. With our residual gauge choice these eight
conditions imply that $h_{ij}(\eta,\vec{x})$ is transverse and traceless at
linearized order. Suppressing gravitons makes it vanish completely at linearized
order.

The only linearized fields which remain are $f$ and $z$. Since $x$ vanishes, they
can depend only upon the operators $Y(\vec{k})$ and $Y^{\dagger}(\vec{k})$,
\begin{eqnarray}
f_{\rm ph}(\eta,{\vec x}) &=& \frac{-1}{\vp_0'(\eta)} \partial_0 \vp_0'(\eta) 
\int {d^3k \over (2\pi)^3} \frac{1}{k} \left\{ Q_C (\eta,k) e^{i {\vec k} 
\cdot {\vec x}} Y({\vec k}) + {\rm h.c.} \right\} \; . \label{eq:fmodes} \\
z_{\rm ph}(\eta,{\vec x}) &=& \frac{\kp}{2} \vp_0'(\eta) \int {d^3k \over 
(2\pi)^3} \frac{1}{k} \left\{ Q_C (\eta,k) e^{i {\vec k} \cdot {\vec x}} Y({\vec k}) 
+ {\rm h.c.} \right\} \; , \label{eq:zmodes}
\end{eqnarray}
We can obviously eliminate $f_{\rm ph}$ in favor of $z_{\rm ph}$,
\begin{equation}
f_{\rm ph}(\eta,\vec{x}) = - \frac2{\kp} {z_{\rm ph}'(\eta,\vec{x}) \over
\vp_0'(\eta)} = {z_{\rm ph}'(\eta,\vec{x}) \over a(\eta) \sqrt{-\dot{H}}} \; .
\end{equation}
This is the well-known constraint between the (dimensionless) 
Newtonian potential $\Phi(\eta,\vec{x}) \equiv \kappa z(\eta,\vec{x})/2a(\eta)$ 
and the scalar field fluctuation \cite{MFB}. 

We can now write down the integrated field equations whose iteration produces
the final perturbative expansion of the various fields,
\begin{eqnarray}
f(\eta,\vec{x}) & = & f_{\rm ph}(\eta,\vec{x}) - \int_{\eta_0}^{\eta} d\eta' 
\int d^3x' \left\{G_{ff}(x;x') {\Delta F}(x') \frac{\mbox{}}{\mbox{}} \right. 
\nonumber \\
& & \qquad \left. \frac{\mbox{}}{\mbox{}} + G_{fz}(x;x') \left[{\Delta E}^{00}(x') 
+ {\Delta E}^{ii}(x') \right]\right\} \; , \label{eq:fphys} \\
z(\eta,\vec{x}) & = & z_{\rm ph}(\eta,\vec{x}) - \int_{\eta_0}^{\eta} d\eta' 
\int d^3x' \left\{G_{zf}(x;x') {\Delta F}(x') \frac{\mbox{}}{\mbox{}} \right. 
\nonumber \\
& & \qquad \left. \frac{\mbox{}}{\mbox{}} + G_{zz}(x;x') \left[{\Delta E}^{00}(x') 
+ {\Delta E}^{ii}(x') \right]\right\} \; , \label{eq:zphys} \\
v_i(\eta,\vec{x}) & = & 2 \int_{\eta_0}^{\eta} d\eta' \int d^3x' G_B(x;x') 
{\Delta E}^{0i}(x') \; , \label{eq:vphys} \\ 
h_{ij}(\eta,\vec{x}) & = & -2 \left(\delta_{ik} \delta_{j\ell} - \delta_{ij} 
\delta_{k\ell}\right) \int_{\eta_0}^{\eta} d\eta' \int d^3x' G_A(x;x') 
{\Delta E}^{k\ell}(x') \; . \qquad \label{eq:hphys} 
\end{eqnarray}
Note that $v_i$ and $h_{ij}$ vanish at $\eta = \eta_0$, along with their first 
derivatives. On the other hand, both $f$ and $z$ suffer perturbative correction 
on the initial value surface. This derives from the existence of a gravitational 
interaction even on the initial value surface, as required by the constraint 
equations of the ungauged formalism.

\subsection{The long wavelength approximation}

Except for suppressing the gravitons, all the results obtained to 
this point have been exact. Unfortunately they are also largely useless 
because we lack explicit forms for the various mode functions and 
retarded Green's functions in a general inflationary background. Nor
could we perform the required integrations if we did possess such 
expressions. What makes calculations possible is the long wavelength 
approximation which effectively removes spatial variation. The 
resulting problem of temporal calculus is still formidable, but it 
can be treated using the slow roll expansion discussed in Section 2.

The back-reaction we seek to study is the response to the superadiabatic
amplification of modes which experience horizon crossing. This is an
intrinsicly infrared phenomenon so one might suspect that it can 
be studied effectively in the limit where spatial derivatives are 
dropped and the mode functions approach their leading infrared forms.
This limit defines the long wavelength approximation and it effects 
three sorts of simplifications on the perturbative apparatus of
relations (\ref{eq:fphys}-\ref{eq:hphys}).

The first simplification is to ignore spatial derivatives in the source 
terms, ${\Delta F}$ and ${\Delta E}^{\mu\nu}$. An immediate consequence of
this, and of the absence of dynamical gravitons, is that ${\Delta E}^{0i}$
vanishes and ${\Delta E}^{ij}$ is proportional its trace,
\begin{equation}
{\Delta E}^{0i} \vert_{\ell.w.} \longrightarrow  0 \quad , \quad 
{\Delta E}^{ij} \vert_{\ell.w.} \longrightarrow \frac13 \delta_{ij} 
{\Delta E}^{kk} \; .
\end{equation}
It follows that $v_i$ remains zero for all time and that only the trace 
part of $h_{ij}$ ever becomes nonzero,
\begin{eqnarray}
v_i(\eta,\vec{x}) \vert_{\ell.w.} & \longrightarrow & 0 \; , \\
h_{ij}(\eta,\vec{x}) \vert_{\ell.w.} & \longrightarrow & \frac43 \delta_{ij} 
\int_{\eta_0}^{\eta} d\eta' \int d^3x' G_A(x;x') {\Delta E}^{kk}(x') \; . 
\end{eqnarray}

The second simplification is that one can replace the mode functions of
the linearized solutions by their infrared limiting forms. These forms
were worked out in \cite{AbWo1} by matching the leading ultraviolet term
of the ultraviolet expansion to the leading term of the infrared expansion
at the time $\eta_*(k)$ of horizon crossing: $k = H(\eta_*) a(\eta_*)$. Since
the physical fields contain only $C$ modes the result we require is,
\begin{equation}
Q_C (\eta,k) \vert_{\ell.w.} \longrightarrow \frac{\sqrt{-\dot{H}(\eta)}}{H^2(\eta)}
\cdot \frac{H^2(\eta_*)}{\sqrt{-\dot{H}(\eta_*)}} {e^{-ik \eta_*} \over
\sqrt{2k}} \; .
\end{equation}
Note also that momentum integrations are cut off at $k = H(\eta) a(\eta)$. 
The physical justification for this is that only modes below the cutoff have 
undergone the superadiabatic amplification that is the basis for the effect 
whose back-reaction we seek to evaluate. However, one must take care that the 
resulting time dependence is a genuine infrared effect and not simply the 
result of allowing more and more modes to contribute to what would be an 
ultraviolet divergence without the cutoff. One consequence is that time
derivatives of the fields should not be allowed to act on the momentum 
cutoff.

The third simplification is that we expand the inverses of 
$\widehat{\cal D}_I \equiv {\cal D}_I + \nabla^2$ in powers of $\nabla^2$,
\begin{equation}
\widehat{\cal D}_I^{-1} = {\cal D}_I^{-1} - {\cal D}_I^{-2} \nabla^2 + \dots 
\end{equation}
Only the zeroth order term is required for $h_{ij}$,
\begin{equation}
h_{ij} \vert_{\ell.w.} \longrightarrow \frac43 \delta_{ij} {\cal D}_A^{-1} 
{\Delta E}^{kk} \; , \label{eq:hlw}
\end{equation}
but one must go to first order for the various $(f,z)$ Green's functions
on account of their prefactors of $1/\nabla^2$. The Appendix demonstrates
that the leading results are,
\begin{eqnarray}
G_{ff}(x;x') \vert_{\ell.w.} & \longrightarrow & \left[ \vp_0' {\cal{D}}_B^{-1} 
\frac{1}{\vp_0'} + 2 \vp_0' {\cal{D}}_B^{-1}  \frac{\vp_0''}{{\vp_0'}^2} 
{\cal{D}}_C^{-1} \vp_0' \partial_0 \frac{1}{\vp_0'} \right] \delta^4(x-x') 
\; , \quad \label{eq:ffir} \\
G_{fz}(x;x') \vert_{\ell.w.} & \longrightarrow & \kappa \vp_0' {\cal{D}}_B^{-1} 
\frac{\vp_0''}{{\vp_0'}^2} {\cal{D}}_C^{-1} \vp_0' \delta^4 (x-x') \; , 
\label{eq:fzir} \\ 
G_{zf}(x;x') \vert_{\ell.w.} & \longrightarrow & \kappa \vp_0' {\cal{D}}_C^{-1} 
\frac{\vp_0''}{{\vp_0'}^2} {\cal{D}}_B^{-1} \vp_0' \delta^4 (x-x') \; ,
\label{eq:zfir} \\
G_{zz}(x;x') \vert_{\ell.w.} & \longrightarrow & \left[ \vp_0' {\cal{D}}_C^{-1}
\frac1{\vp_0'} + 2 \vp_0' {\cal{D}}_C^{-1} \frac{\vp_0''}{{\vp_0'}^2} 
{\cal{D}}_B^{-1} \partial_0 \right] \delta^4(x-x') \; . \quad \label{eq:zzir}
\end{eqnarray}

At this point we can make contact with the similar calculation that 
was done in \cite{AbWo1}. In the infrared limit, the general retarded 
propagators (\ref{eq:ffir})-(\ref{eq:zzir}) correspond exactly to the
result that can be inferred from Eqs. (131)-(132) of \cite{AbWo1}.
The ``$C$'' and ``$D$'' in Eqs. (131) and (132) from that paper (where 
$\Omega \equiv a$)  correspond respectively to the change induced in the 
$z$ and in the $f$ fields that obtains from our equations
(\ref{eq:fphys}-\ref{eq:zphys}). The the amputated 1-point functions in
those equations correspond to our source terms according to the rule: 
$\tilde{\alpha} \rightarrow -\frac{\kp}3 a {\Delta E}^{kk}$, $\tilde{\gamma} 
\rightarrow - \kp a {\Delta E}^{00}$ and $\tilde{\delta} \rightarrow -\kp a
{\Delta F}$.

\section{Quadratic corrections}

The purpose of this section is to apply the technology of Section 3 to
obtain the scalar and metric at quadratic order in the initial value 
operators $Y(\vec{k})$ and $Y^{\dagger}(\vec{k})$, and to leading order 
in the slow roll and long wavelength expansions. In Ref. \cite{AbWo1} we 
computed the one loop expectation values of the scalar and metric fields, 
also to leading order in the slow roll and long wavelength expansions. Much 
of that calculation can be used to get the quadratic order operator 
expansions of the same quantities. In particular, all the vertices are 
catalogued in that reference.

As an example, consider vertex \# 9 of Table 3 from \cite{AbWo1}. That 
vertex corresponds to the following cubic term in the Lagrangian,
\begin{equation}
{\cal L}_{\#9} = -\frac{\kp}4 V_{,\vp\vp}(\vp_0) \, a^4 \phi^2 \psi =
-\frac{\kp}4 V_{,\vp\vp}(\vp_0) \, a f^2 (2z + h) \; . \label{eq:v9}
\end{equation}
Variation with respect to $\phi$ and $\psi_{\mu\nu}$ gives the following 
source terms,
\begin{equation}
{\Delta F}_{\#9} = -\kp V_{,\vp\vp}(\vp_0) \, a f (z + \frac12 h) \quad , 
\quad {\Delta E}^{\mu\nu}_{\#9} = -\frac{\kp}4 V_{,\vp\vp}(\vp_0) \, a f^2
\eta^{\mu\nu} \; .
\end{equation}
The lowest order perturbative corrections from these terms come when 
$f$, $z$ and $h$ are replaced by the associated physical linearized 
fields: $f_{\rm ph}$, $z_{\rm ph}$ and $0$, respectively. Recall also
that $f_{\rm ph} = z'/a\sqrt{-\dot{H}}$. 

The preceding are all exact results. Because we are also making the long 
wavelength and slow roll approximations, $z_{\rm ph}$ can be written as,
\begin{equation}
z_{\rm ph}(\eta,\vec{x}) \vert_{\ell.w.+s.r.} \longrightarrow a(\eta)
\left[{-\dot{H}(\eta) \over H^2(\eta)}\right] \int {d^3k 
\over (2\pi)^3} {H_* \over \sqrt{2 k^3}} \sqrt{H^2_* \over - \dot{H}_*}
\left\{ Y(\vec{k}) + Y^{\dagger}(\vec{k}) \right\} \; . \label{eq:zY2}
\end{equation}
To leading order in the slow roll approximation the time depedence of
this expression comes entirely in the initial factor of $a(\eta)$,
which allows the following simplification for $f_{\rm ph}$,
\begin{equation}
f_{\rm ph}(\eta,\vec{x}) \vert_{\ell.w.+s.r} \longrightarrow \sqrt{
H^2(\eta) \over -\dot{H}(\eta)} z_{\rm ph}(\eta,\vec{x}) \; .
\end{equation}
We can also use the slow roll approximation to evaluate derivatives
of the scalar potential,
\begin{equation}
\kp V_{,\vp}(\vp_0) \vert_{s.r.} \longrightarrow 6 H \sqrt{ -\dot{H}} 
\quad \quad , \quad \quad V_{,\vp \vp}(\vp_0) \vert_{s.r.} \longrightarrow
-3 \dot{H} \; .  \label{backid}
\end{equation}
The source terms finally contributed by vertex \#9 in the slow roll 
and long wavelength approximations are,
\begin{eqnarray}
{\Delta F}_{\# 9} \vert_{\ell.w. + s.r.} & \longrightarrow & -3 \kappa
H^2 \sqrt{-\dot{H} \over H^2} a z_{\rm ph}^2 \; , \\
{\Delta E}^{00}_{\# 9} \vert_{\ell.w. + s.r.} & \longrightarrow & +\frac34
\kappa H^2 a z_{\rm ph}^2 \; , \\
{\Delta E}^{kk}_{\# 9} \vert_{\ell.w. + s.r.} & \longrightarrow & -\frac94
\kappa H^2 a z_{\rm ph}^2 \; .
\end{eqnarray}
Summing the contributions from all vertices of Einstein-scalar 
\cite{AbWo1} under the same set of approximations gives the following
quadratic source terms,
\begin{eqnarray}
{\Delta F} \vert_{\ell.w. + s.r.} & \longrightarrow & 3 \kappa
H^2 \sqrt{-\dot{H} \over H^2} a z_{\rm ph}^2 \; , \\
{\Delta E}^{00} \vert_{\ell.w. + s.r.} & \longrightarrow & -\frac94
\kappa H^2 a z_{\rm ph}^2 \; , \\
{\Delta E}^{kk} \vert_{\ell.w. + s.r.} & \longrightarrow & +\frac{27}4
\kappa H^2 a z_{\rm ph}^2 \; . \label{eq:hsource}
\end{eqnarray}

The next step is to apply the various retarded Green's functions in the
long wavelength and slow roll approximations. As an example, we substitute
(\ref{eq:hsource}) into (\ref{eq:hlw}) to obtain the leading quadratic 
correction to $h_{ij}$,
\begin{equation}
h^{(2)}_{ij} \vert_{\ell.w.+s.r.} \longrightarrow - 9 \kp \delta_{ij} 
a \int^\eta_{\eta_0} d \eta' \, a^{-2} \int^{\eta'}_{\eta_0} d\eta'' \, 
a^2 H^2 z_{\rm ph}^2 \; , 
\end{equation}
where $z_{\rm ph}$ is given by Eq. (\ref{eq:zY2}). Taking the momentum 
integrals outside, we are left with the following integrations over time:
\begin{equation}
a(t)\int^t_{t_0} dt' a^{-3} (t') \int^{t'}_{t_0} dt'' a^3(t'') 
\frac{\dot{H}^2}{H^2} = \frac16 a (t) \left(\frac{-\dot{H}}{H^2}\right)
+ \ldots 
\end{equation}
The result is therefore,
\begin{equation}
h_{ij} \vert_{\ell.w.+s.r.} \longrightarrow \delta_{ij} \left({H^2 \over 
- \dot{H}}\right) \left\{ -\frac32 {\kp z_{\rm ph}^2 \over a} + 
O(z_{\rm ph}^3)\right\} \; .
\end{equation}
The analogous reductions for $f$ and $z$ are straightforward but rather
tedious, owing to their complicated mixing. The final answer is,
\begin{eqnarray}
f \vert_{\ell.w.+s.r.} & \longrightarrow & \sqrt{H^2 \over \dot{H}} 
\left\{ z_{\rm ph} + \frac54 {\kp z_{\rm ph}^2 \over a} + 
O(z_{\rm ph}^3)\right\} \; , \\
z \vert_{\ell.w.+s.r.} & \longrightarrow & \left\{z_{\rm ph} + 
\frac72 {\kp z_{\rm ph}^2 \over a} + O(z_{\rm ph}^3)\right\} \; .
\end{eqnarray}

We invert (\ref{eq:vars}) to recover $\phi$ and $\psi_{\mu\nu}$.
It is useful to multiply by $\kappa$ to absorb the dimensions,
and to express the result in terms of the (dimensionless) Newtonian 
potential:
\begin{equation}
\Phi(x) \equiv \frac{\kp z_{\rm ph}(x)}{2 a(\eta)} \; . \label{newt}
\end{equation}
Note that, from Eqs. (\ref{newt}) and (\ref{eq:zY2}) we have $\Phi 
\propto H^{-2}$ (since $\dot{H} \sim$ constant) so that its time 
derivative is down from $H \Phi$ by a slow roll parameter,
\begin{equation}
\dot{\Phi} \vert_{\ell.w.+s.r.} \longrightarrow 2 \left({-\dot{H} \over 
H^2}\right)  H \Phi \; . \label{dotnewt}
\end{equation}
With this terminology, the nonzero perturbed fields are,
\begin{eqnarray}
\kp \phi \vert_{\ell.w.+s.r.} & \longrightarrow & \sqrt{H^2 \over -\dot{H}} 
\left\{ 2 \Phi + 5 \Phi^2 + O(\Phi^3) \right\} \; . \label{fi} \\
\kp \psi_{00} \vert_{\ell.w.s.r.} & \longrightarrow & 2\Phi + 14 \Phi^2 + 
O(\Phi^3) \; , \label{psi00} \\ 
\kp \psi_{ij} \vert_{\ell.w.+s.r.} & \longrightarrow & \delta_{ij} 
\left\{ 2\Phi - 6 \left(\frac{H^2}{-\dot{H}}\right) \Phi^2 + O(\Phi^3)
\right\} \; . 
\label{psiij}
\end{eqnarray}

\section{An invariant measure of expansion}

The purpose of this section is to compute corrections through quadratic
order to the operator ${\cal A}_{\rm inv}$ we have proposed \cite{Obs1} 
as an invariant measure of the rate of cosmological expansion. We first
expand the scalar ${\cal A} \equiv \Box_c^{-1}$ in powers of the
pseudo-graviton field and then substitute (\ref{psi00}-\ref{psiij})
to express the result as a function of the unconstrained initial
value operators, to leading order in the slow roll and long wavelength
approximations. Full invariance is achieved by evaluating the ${\cal A}(
\eta,\vec{x})$ in a geometrically specified coordinate system.

\subsection{The scalar observable}

The pseudo-graviton expansion is most easily accomplished by first expressing 
$\Box_c$ in terms of the conformally rescaled metric,
\begin{equation}
\widetilde{g}_{\mu\nu}(\eta,\vec{x}) \equiv a^{-2}(\eta) g_{\mu\nu}(\eta,\vec{
x}) = \eta_{\mu\nu} + \kappa \psi_{\mu\nu}(\eta,\vec{x}) \; .
\end{equation}
We write ${\Box}_c = a^{-3} {\cal D} a$, where ${\cal D}$ and its expansion
are,
\begin{equation}
{\cal D} \equiv {1 \over \sqrt{-\widetilde{g}}} \partial_{\mu} \left(\sqrt{-
\widetilde{g}} \: \widetilde{g}^{\mu\nu} \partial_{\nu}\right) - \frac16
\widetilde{
R} = \partial^2 + \kappa {\cal D}_1 + \kappa^2 {\cal D}_2 + \dots \; .
\end{equation}
The first two operators in the expansion are,
\begin{eqnarray}
{\cal D}_1 &=& - \psi^{\mu\nu} \partial_\mu \partial_\nu + \left(-\psi^{\mu
\alpha}_{~~, \alpha} + \frac12 \psi^{,\mu} \right) \partial_{\mu} - \frac16
\left( \psi^{\rho\sigma}_{~~ , \rho\sigma} - \psi^{,\rho}_{~~\rho}\right) \; ,
\label{eq:D1} \\
{\cal D}_2 &=& \psi^{\mu\alpha} \psi^{\nu}_{\; \alpha} \partial_{\mu}
\partial_{\nu} + \left( (\psi^{\alpha\beta} \psi^{\mu}_{~\alpha})_{, \beta} -
\frac12 \psi^{\alpha\beta,\mu} \psi_{\alpha\beta} + \frac12 \psi^{\alpha\mu}
\psi_{,\alpha}\right) \partial_{\mu} -\frac16 \widetilde{R}_2 . \qquad
\label{D2}
\end{eqnarray}
The second order, conformally rescaled Ricci scalar is,
\begin{eqnarray}
\lefteqn{\widetilde{R}_2 \equiv \psi^{\alpha\beta} \left(\psi^{~~~ ,\gamma}_{
\alpha\beta ~~ \gamma} + \psi_{,\alpha\beta} - 2 \psi^{\gamma}_{~ \alpha ,
\beta \gamma} \right) } \nonumber \\
& & + \frac34 \psi^{\alpha\beta,\gamma} \psi_{\alpha\beta,\gamma} - \frac12
\psi^{\alpha \beta ,\gamma} \psi_{\gamma \beta,\alpha} - \psi^{\alpha \beta}_{
~~~ ,\beta} \psi^{\gamma}_{~ \alpha , \gamma} +\psi^{\alpha \beta}_{~~~ ,\beta}
\psi_{, \alpha} - \frac14 \psi^{,\alpha} \psi_{, \alpha} \; . \quad
\end{eqnarray}

The next step is to factor $\partial^2$ out of ${\cal D}$,
\begin{equation}
{\cal D} = \partial^2 \left(1 + \frac1{\partial^2} \kappa {\cal D}_1 + \frac1{
\partial^2} \kappa^2 {\cal D}_2 + O(\kappa^3)\right) \; .
\end{equation}
Inverting ${\cal D}$ is now straightforward,
\begin{equation}
\frac1{\cal D} = \frac{1}{\partial^2} - \frac{1}{\partial^2} \kappa {\cal
D}_1 \frac{1}{\partial^2} + \frac1{\partial^2} \kappa {\cal D}_1 \frac1{
\partial^2} \kappa {\cal D}_1 \frac{1}{\partial^2} - \frac{1}{\partial^2}
\kappa^2 {\cal D}_2 \frac{1}{\partial^2} + O(\kappa^3) \; .
\end{equation}
All this implies the following expansion for the scalar observable,
\begin{equation}
{\cal A}[g] = a^{-1} \frac1{\cal D} a^3 = {\cal A}_0 + \kappa {\cal A}_1 +
\kappa^2 {\cal A}_2 + O(\kappa^3) \; , \label{A_exp}
\end{equation}
where the first two corrections are,
\begin{eqnarray}
{\cal A}_1 & \equiv & - a^{-1} \frac1{\partial^2} {\cal D}_1 \frac1{\partial^2}
a^3 \; , \\
{\cal A}_2 & \equiv & - a^{-1} \frac1{\partial^2} {\cal D}_2 \frac1{\partial^2}
a^3 + a^{-1} \frac1{\partial^2} {\cal D}_1 \frac1{\partial^2} {\cal D}_1
\frac1{\partial^2} a^3 \; .
\end{eqnarray}

The zeroth-order term ${\cal A}_0 \equiv a^{-1} \partial^{-2} a^3$ has a very 
simple expression in terms of the background expansion rate,
\begin{equation}
{\cal{A}}_0 =  - a^{-1} \int^\eta_{\eta_0} d\eta' \, \int^{\eta'}_{\eta_0} 
d\eta'' a^3(\eta'') = {-1 \over 2 H^2} \left\{1 + O\left({ -\dot{H} 
\over H^2}\right) \right\} \; .
\end{equation}
Our proposal is to define the general expansion rate, as an operator,
to bear the same relation to the full scalar ${\cal A}$ \cite{Obs1}.
Astronomers do the same thing when they infer the Hubble constant under
the assumption that the relation between luminosity distance and redshift
is the same in the actual universe as for a perfectly homogeneous and
isotropic one.

The higher order contributions ${\cal{A}}_1$ and ${\cal{A}}_2$ are
nominally inhomogeneous but lose their dependence upon space in the
long wavelength approximation. The consequent suppression of spatial
derivatives and anisotropic components of the pseudo-graviton field
reduces ${\cal D}_1$ to the following simple form,
\begin{equation}
{\cal D}_1 \vert_{\ell.w.} \longrightarrow -\psi_{00} \partial_0^2
-\frac12 (\psi_{00}' + \psi_{ii}') \partial_0 - \frac16 \psi_{ii}'
\; .
\end{equation}
Further simplifications result from the fact that (co-moving) time
derivatives of $\Phi$ are weaker than $H \Phi$ by a slow roll
parameter as per equation (\ref{dotnewt}). Only the quadratic
term of $\psi_{ij}$ possesses the enhancement necessary to survive
differentiation at leading order,
\begin{eqnarray}
\kappa \psi_{00}' \vert_{\ell.w.+s.r.} & \longrightarrow & H a
\left\{0 + O(\Phi^3)\right\} \; , \\
\kappa \psi_{ij}' \vert_{\ell.w.+s.r.} & \longrightarrow & H a
\left\{-12 \Phi^2 + O(\Phi^3)\right\} \delta_{ij} \; .
\end{eqnarray}
Of course subsequent conformal time derivatives are dominated by the
factor(s) of $a$,
\begin{eqnarray}
\kappa \psi_{00}'' \vert_{\ell.w.+s.r.} & \longrightarrow & H^2 a^2 
\left\{0 + O(\Phi^3)\right\} \; , \\
\kappa \psi_{ij}'' \vert_{\ell.w.+s.r.} & \longrightarrow & H^2 a^2
\left\{-12 \Phi^2 + O(\Phi^3)\right\} \delta_{ij} \; .
\end{eqnarray}
The result is that only a few terms in ${\cal D}_1$ and ${\cal D}_2$
can contribute at leading order,
\begin{eqnarray}
\kappa {\cal D}_1 \vert_{\ell.w.+s.r.}& \longrightarrow & - 2 \Phi
\partial_0^2 + \Phi^2 \left\{-14 \partial_0^2 + 18 H a \partial_0
+ 6 H^2 a^2\right\} \; , \\
\kappa {\cal D}_2 \vert_{\ell.w.+s.r.}& \longrightarrow & -4 \Phi^2 
\partial_0^2 + O(\Phi^3) \; .
\end{eqnarray}

Because $\dot{\Phi} \ll H \Phi$ we can ignore its time dependence with
respect to factors of $a(\eta)$. This makes acting the various factors
of $\partial^{-2}$ quite simple in the slow roll approximation,
\begin{eqnarray}
{\cal A}_1 \vert_{\ell.w.+s.r.} & \longrightarrow & \frac1{a} \frac1{
\partial^2} a^3 \left(-2 \Phi - 2 \Phi^2 + O(\Phi^3)\right) \; , \\
& \longrightarrow & {-1 \over 2 H^2} \left\{-2 \Phi - 2 \Phi^2 +
O(\Phi^3)\right\} \; .
\end{eqnarray}
Since the two terms in ${\cal A}_2$ cancel to leading order, the 
result is,
\begin{equation}
{\cal{A}}(x) \vert_{\ell.w.+s.r.} \longrightarrow {-1 \over 2 H^2} 
\left\{ 1 - 2 \Phi - 2 \Phi^2 + O(\Phi^3) \right\} \; . \label{APhi}
\end{equation}

\subsection{The invariant observable}

Even a scalar changes under coordinate transformation. To achieve 
full invariance we measure ${\cal A}[g](x)$ at a point $Y^{\mu}[\vp,g](x)$
which is invariantly specified in terms of the operators $\vp$ and 
$g_{\mu\nu}$. Since there is no spatial dependence in the long wavelength 
approximation we actually need only fix the 3-surface, $Y^0 = \tau$. 
Following the suggestion of Unruh, we chose the functional 
$\tau[\vp](\eta,\vec{x})$ to make the full scalar agree with its 
background value at conformal time $\eta$,
\begin{equation}
\vp\left({\mbox{} \over \mbox{}} \tau(\eta,\vec{x}),\vec{x}\right) =
\vp_0(\eta) \; .
\end{equation}
The weak field expansion of $\tau[\vp]$ is \cite{Obs1},
\begin{equation}
\tau[\vp](\eta,\vec{x}) = \eta - {\phi(\eta,\vec{x}) \over \vp_0'(\eta)}
+ {\phi(\eta,\vec{x}) \phi'(\eta,\vec{x}) \over \vp_0^{\prime 2}(\eta)}
-{ \vp_0''(\eta) \over 2 \vp_0'(\eta)} \left({\phi(\eta,\vec{x}) \over 
\vp_0'(\eta)} \right)^2 +O(\phi^3) \; .
\end{equation}
Substitution of our result (\ref{fi}) for the perturbed scalar gives the 
final expansion to leading order in the slow roll and long wavelength
approximations,
\begin{equation}
\tau \vert_{\ell.w.+s.r} \longrightarrow \eta + \left({H^2 \over 
-\dot{H}}\right) {\Phi \over H a} - \left[\left({H^2 \over -\dot{H}}
\right)^2 - 7 \left({H^2 \over -\dot{H}}\right)\right] {\Phi^2 \over
2 H a} + O(\Phi^3) \; . \label{eq:tau}
\end{equation}

Our invariant expansion operator is just ${\cal A}$ evaluated at this
point,
\begin{equation}
{\cal A}_{\rm inv}[\vp,g](\eta,\vec{x}) \equiv {\cal A}[g]\left(
{\mbox{} \over \mbox{}} \tau[\vp](\eta,\vec{x}),\vec{x}\right) \; .
\end{equation}
Note that there is no obstacle in perturbation theory to evaluating
an operator at a point $\tau[\vp] \equiv \eta + {\delta \tau}[\vp]$
which is itself an operator,
\begin{equation}
{\cal A}_{\rm inv} \equiv {\cal A} + {\cal A}' {\delta \tau} + \frac12
{\cal A}'' {\delta \tau}^2 + \dots 
\end{equation}
The derivatives are straightforward to evaluate with (\ref{APhi}) and
(\ref{dotnewt}),
\begin{eqnarray}
{\cal A}' \vert_{\ell.w.+s.r.} & \longrightarrow & {-H a \over 2 H^2}
\left({-\dot{H} \over H^2}\right) \left\{2 - 8 \Phi + O(\Phi^2)\right\}
\; \\
{\cal A}'' \vert_{\ell.w.+s.r.} & \longrightarrow & {-H^2 a^2 \over 2 H^2}
\left\{2 \left({-\dot{H} \over H^2}\right) + 6 \left({-\dot{H} \over H^2}
\right)^2 + O(\Phi)\right\} \; .
\end{eqnarray}
However, putting everything together results in complete cancellation of
all corrections to the order we are working,
\begin{equation}
{\cal{A}}_{\rm inv} = {-1 \over 2 H^2} \left\{1 + {\cal{O}}(\Phi^3)
\right\} \; . \label{RESULT}
\end{equation}

Although this result surprised us, it could have been anticipated
by noting that, in the slow roll approximation, our scalar degenerates 
to a local algebraic function of the Ricci scalar \cite{Obs1},
\begin{equation}
{\cal A}[g](x) \vert_{s.r.} \longrightarrow {-6 \over R(x)} \; .
\end{equation}
Einstein's equations completely determine the Ricci scalar, as an 
operator, from the local matter stress tensor,
\begin{eqnarray}
R(x) & = & - 8\pi G g^{\mu\nu} T_{\mu\nu} \; , \\
& = & 16 \pi G \left\{ V(\vp) - \frac12 g^{\mu\nu} \partial_{\mu} \vp
\partial_{\nu} \vp \right\} \; .
\end{eqnarray}
We can always choose to work in a coordinate system for which the
full scalar agrees with its background value. When this is done one
sees that back-reaction can only enter through the kinetic term. Since
it is certainly from kinetic effects that the pure gravitational result
derives \cite{TW1}, there may well be a significant back-reaction from
the scalar kinetic term as well. Unfortunately, one can never tell 
whether there is or not when the kinetic term is systematically 
neglected, which is just what happens when the slow roll and long 
wavelength approximations are combined. It follows that our 
approximations must produce a null result --- whether or not there 
really is significant back-reaction --- not just at quadratic order 
in the initial value operators, but at all higher orders as well.

\section{Stochastic samples}

The last section has demonstrated that scalar-driven inflation can show 
no secular back-reaction to leading order in the long wavelength and slow 
roll approximations. Since this holds as a strong operator equation for 
our expansion observable, ${\cal A}_{\rm inv}$, there is no need to choose 
between expectation values and stochastic samples. However, we believe
the case is still quite strong for secular back-reaction when the long
wavelength approximation is relaxed, although it would have to come from 
the coherent superposition of interactions at higher than quadratic order
in the initial value operators. The purpose of this section is to develop 
a theoretical framework for studying such an effect through stochastic 
samples as recommended by Linde and others \cite{Pritzker}. We begin by
motivating and defining the basic formalism, then we treat the crucial
question of the degree of stochastic fluctuation expected in functionals 
of the fields such as ${\cal A}_{\rm inv}[\vp,g]$.

\subsection{Motivation and basic formalism}

The simplest way to motivate stochastic effects is by considering the
linearly independent mode functions of the infrared regime, $Q_{1,I}(\eta,k)$ 
and $Q_{2,I}(\eta,k)$, which were described in subsection 3.2. The {\it 
full} linearized field operator must involve both of these mode functions
in order to avoid commuting with its conjugate momentum. However, after 
horizon crossing ($k = H a$) one of the mode functions becomes vastly 
larger than the other. This is the phenomenon of superadiabatic amplification 
and it corresponds to the simple physical picture of particle production
through infrared virtual quanta becoming trapped in the inflationary 
Hubble flow. If one retains just the larger mode function then the linearized
operator which results is effectively classical in that it commutes with 
its time derivative. Although the result is still probabilistic, one can
simultaneously measure the value of such an operator and its time 
derivative, and subsequent measurements will show only the classical 
evolution expected from these initial values. This is the reflection, in
the Heisenberg picture, of a ``squeezed state.''

The expectation value of a functional of squeezed operators can fail to 
provide a good estimate of what an actual observer sees. For example, the 
expectation value of the current stress tensor is presumably homogeneous 
and isotropic. We see nothing of the sort because we exist at the end
of a long period of essentially classical evolution from one particular
choice, random but definite, of superadiabatically amplified density 
perturbations.

A better way of treating squeezed operators is to sample the result of
making random but definite choices for them from the relevant quantum
mechanical wave function, and then evolving classically. It is crucial 
to understand that taking such a stochastic sample is perfectly consistent 
with the use of quantum field theory to express the Heisenberg operators
as functionals of the unconstrained initial value operators. Nor is there
any change in how the observable ${\cal A}_{\rm inv}(\eta,\vec{x})$ 
depends upon the Heisenberg operators. (Of course we do want to avoid 
making the long wavelength approximation!) What changes is that the 
scalar creation and annihilation operators --- $Y^{\dagger}(\vec{k})$ and 
$Y(\vec{k})$ --- are random numbers up to $k = H(\eta) a(\eta)$, and zero 
beyond.

To avoid problems with continuum normalization we take the 3-manifold to 
be $T^3$ with identical co-moving coordinate radii of $H_0^{-1} \equiv
H^{-1}(\eta_0)$. (Because the conformal coordinate volume is so restricted 
during inflation the integral approximation is excellent for mode sums and 
there is no conflict with any of our previous, continuously normalized, 
results.) On this manifold the co-moving wavevectors become discrete,
\begin{equation}
\vec{k} = 2 \pi H_0 \vec{n} \; .
\end{equation}
Phase space integrals are converted into mode sums in the usual way,
\begin{equation}
\int {d^3 k \over (2\pi)^3} f(\vec{k})  \longrightarrow H_0^3 \sum_{\vec{n}} 
f(2\pi H_0 \vec{n}) \; .
\end{equation}
And the Dirac delta function goes into a Kronecker one,
\begin{equation}
(2 \pi)^3 \delta^3(\vec{k}-\vec{k}') \longrightarrow H_0^{-3} 
\delta_{\vec{n} ,\vec{n}'} \; .
\end{equation}

Because the initial scalar state is free, the associated creation and
annihilation operators are stochastically realized as independent,
complex, Gaussian random variables with standard deviation $H_0^{-3}$.
It is convenient to scale out the dimensions,
\begin{equation}
A_{\vec{n}} \equiv H_0^{\frac32} Y(2\pi H_0 \vec{n}) \quad , \quad
A^*_{\vec{n}} \equiv H_0^{\frac32} Y^{\dagger}(2\pi H_0 \vec{n}) \; ,
\end{equation}
so that the probability density for each mode $\vec{n}$ has a simple 
expression,
\begin{equation}
\rho(A_{\vec{n}},A^*_{\vec{n}}) = {1 \over 2\pi} e^{-A_{\vec{n}} 
A^*_{\vec{n}}} \; . \label{eq:gauss}
\end{equation}
Since the modes are independent, the joint probability distribution
is just the product of (\ref{eq:gauss}) over the relevant range of
$\vec{n}$.

\subsection{Functionals of stochastic variables}

Our observable ${\cal A}_{\rm inv}(\eta,\vec{x})$ depends in a 
complicated way upon the scalar and metric fields, which are themselves
functionals of the stochastic variables $A_{\vec{n}}$ and $A^*_{\vec{n}}$. 
The nature of this dependence determines the crucial issues of 
whether or not ${\cal A}_{\rm inv}(\eta,\vec{x})$ is well represented 
by its expectation value and whether or not a definite sign can be
inferred for corrections to the cosmological expansion rate. Of course 
we do not yet have a replacement for the long wavelength approximation
which shows secular back-reaction. However, we do here develop a
technique for separating nonlinear functionals of the stochastic 
variables into a part whose percentage fluctuation becomes negligible 
for a long period of inflation, plus another part whose fluctuation is
not negligible but which has a definite sign.

Recall from probability theory that a functional $f = F[A,A^*]$ of 
random numbers is itself a random number. Its probability distribution 
function descends from that of the $A_{\vec{n}}$ and $A^*_{\vec{n}}$ by 
Fourier transformation,
\begin{eqnarray}
\rho(f) & = & \int_{-\infty}^{\infty} {dk \over 2\pi} e^{i k f} 
\left\langle e^{-i k F[A,A^*]} \right\rangle \; , \\
& = & \int_{-\infty}^{\infty} {dk \over 2\pi} e^{i k f} \left(
\prod_{\vec{n}} \int {dA_{\vec{n}} dA^*_{\vec{n}} \over 2\pi} 
e^{-A_{\vec{n}} A^*_{\vec{n}}} \right) e^{-i k F[A,A^*]} \; . 
\label{probdist}
\end{eqnarray}
As an example, consider the Newtonian potential ({\ref{newt}),
\begin{equation}
\Phi(\eta,\vec{x}) = {\kappa \sqrt{-H_0 \dot{H}(\eta)} \over 4\pi} 
\sum_{\vec{n}} \frac1{n} \left\{Q_C(\eta,2\pi H_0 n) e^{i2\pi H_0 {\vec n} 
\cdot {\vec x}} A_{\vec{n}} + {\rm c.c.}\right\} \; .
\end{equation}
The various integrations are trivial Gaussians, as they would be for
any linear variable. The result is that $\Phi$ follows a Gaussian 
distribution with mean zero,
\begin{equation}
\rho(\Phi) = {1 \over \sqrt{2\pi \sigma^2}} e^{-\Phi^2/2\sigma^2} \; ,
\end{equation}
and a spatially constant variance equal to,
\begin{equation}
\sigma^2(\eta) = {-\kappa^2 H_0 \dot{H} \over 8\pi^2} \sum_{\vec{n}} 
\frac1{n^2} \Vert Q_C(\eta,2\pi H_0 n) \Vert^2 \; .
\end{equation}
Making the slow roll and long wavelength approximations we see that
the variance grows as the logarithm of the scale factor,
\begin{eqnarray}
\sigma^2(\eta)\vert_{\ell.w.+s.r.} & \longrightarrow & {\kappa^2 
\over 32 \pi^3} \left({-\dot{H} \over H^2} \right)^2 \sum_{\vec{n}} 
{H_*^2 \over n^3} \left({H_*^2 \over -\dot{H}_*}\right) \; , \\
& \longrightarrow & {\kappa^2 \over 8 \pi^2} \left({-\dot{H} \over H^2} 
\right)^2 \int_1^a {dn \over n} \left({H_*^4 \over -\dot{H}_*} \right) \; , \\
& \longrightarrow & {\kappa^2 H^2 \over 8 \pi^2} \left({-\dot{H} \over H^2} 
\right) \ln[a(\eta)] \; . \label{eq:varest}
\end{eqnarray}

The simplest sort of nonlinear functional is just the square of a linear
one. Since linear functionals of $A_{\vec{n}}$ and $A^*_{\vec{n}}$ are
always Gaussian, their squares always follow a Chi-squared distribution 
whose mean is the variance of the Gaussian and whose variance is twice 
the square of this. For example, the variable $\Phi^2(\eta,\vec{x})$ 
follows a Chi-squared with mean $\sigma^2$ and standard deviation $\sqrt{2} 
\sigma^2$. Although the fluctuations of $\Phi^2$ are of the same order as 
its mean, the sign is definite. Note also that, for a long period of 
inflation, only an incredibly fortuitous sequence of choices for the 
stochastic variables $A_{\vec{n}}$ and $A^*_{\vec{n}}$ would result in 
$\Phi^2$ having a value significantly below some constant times our 
estimate (\ref{eq:varest}). So that if some reliable approximation scheme
should wind up giving the effective expansion rate as,
\begin{equation}
H_{\rm eff}(\eta,\vec{x}) = H(\eta) \left\{1 + \Phi(\eta,\vec{x}) - 
\Phi^2(\eta,\vec{x}) \right\} \; ,
\end{equation}
then the conclusion would be that there is only a vanishingly small 
probability to observe anything except a secular slowing of inflation.
This is an example of how stochastic samples might show the same 
qualitative results as expectation values while breaking exact 
homogeneity and isotropy and altering the numerical coefficient of
the order of magnitude estimate.

It might be thought that quantitative results are not obtainable for
more complicated nonlinear functionals. That useful statements can still 
be made derives from the following fact: {\it any secular back-reaction 
effect must result from the coherent superposition of contributions from 
an enormously large number of modes.} This allows us to exploit the same
sorts of simplifications that underlie statistical mechanics. 

To illustrate the important considerations without becoming too mired in
technical detail let us consider quadratic superpositions of the form,
\begin{equation}
s = S[A,A^*] \equiv \frac12 \sum_{\vec{m}, \vec{n}} S_{\vec{m} \vec{n}} 
\left(A_{\vec{m}} + A^*_{\vec{m}}\right) \left(A_{\vec{n}} + 
A^*_{\vec{n}}\right) \; .
\end{equation}
The characteristic function is,
\begin{equation}
\left\langle e^{-i k S[A,A^*]} \right\rangle = {1 \over \sqrt{{\rm det}(
I + 2 i k S)}} \; ,
\end{equation}
where $S$ stands for the symmetric matrix $S_{\vec{m} \vec{n}}$ and $I$ is 
the unit matrix of the same rank. The various moments of $s$ follow by 
differentiation,
\begin{eqnarray}
\langle s \rangle = {\rm Tr}[S] & , & \left\langle (s - {\rm Tr}[S])^2 
\right\rangle = 2 {\rm Tr}[S^2] \quad , \\
\left\langle (s - {\rm Tr}[S])^3 \right\rangle = 4 {\rm Tr}[S^3] & , &
\qquad \ldots 
\end{eqnarray}

How much fluctuation one should expect is governed by the relation between
traces of powers of the matrix $S$. We distinguish two cases:
\begin{enumerate}
\item ``Local'' superpositions which are characterized by ${\rm Tr}[S^n] 
= ({\rm Tr}[S])^n$ for $n \ge 2$; and
\item ``Nonlocal'' superpositions which obey ${\rm Tr}[S^2] \ll ({\rm 
Tr}[S])^2$ and ${\rm Tr}[S^n] \ll {\rm Tr}[S^2] ({\rm Tr}[S])^{n-2}$ for
$n \ge 3$.
\end{enumerate}
The local case is just the square of a linear superposition and has 
already been considered. To recapitulate, it has significant fluctuation
but definite sign. The distribution for a nonlocal superposition can be
approximated by dropping higher traces of $S$ in the exponent of 
(\ref{probdist}),
\begin{eqnarray}
\rho(s) & = & \int_{-\infty}^{\infty} {dk \over 2\pi} e^{i k s - \frac12
{\rm Tr}[\ln(I + 2 i k S)]} \; , \\
& \approx & \int_{-\infty}^{\infty} {dk \over 2\pi} e^{i k s - i k {\rm
Tr}[S] - k^2 {\rm Tr}[S^2]} \; , \\
& = & {1 \over \sqrt{4\pi {\rm Tr}[S^2]}} \exp\left[- {(s - {\rm Tr}[S])^2 
\over 4 {\rm Tr}[S^2]}\right] \; .
\end{eqnarray}
This is just a Gaussian centered on ${\rm Tr}[S]$ with variance $2 Tr[S^2]$. 
Since its standard deviation is insignificant compared with the mean, 
stochastic effects are not important.

Recall that invariant measures of the cosmological expansion rate derive
their dependence upon $A_{\vec{n}}$ and $A^*_{\vec{n}}$ from two 
expansions. In the first the observable is expanded in powers of the
perturbed fields, then one substitutes the expansions of the perturbed 
fields in powers of the stochastic initial value data. Both of these
expansions really involve averages over the past lightcone of the 
observation point. However, depending upon the observable and the
approximation techniques used to solve for the fields, the terrific
expansion of spacetime during inflation may weight the averages 
heavily towards the most recent times. In that case one gets a local
superposition and stochastic effects are important but simple to
include. An improved observable can also be defined so that the entire 
past lightcone participates effectively \cite{Obs1}, and it seems 
likely that improving on the long wavelength approximation will have
this result anyway. In this case one tends to get a nonlocal 
superposition because modes with different $\vec{n}$ interfere 
destructively. The stochastic fluctuation of back-reaction would
then be negligible and one may as well resort to expectation values.

There is also the possibility of mixing of local and nonlocal, in
which case the best strategy is to separate the two effects and
treat them as above. One might anticipate that analytic considerations
would render the local part obvious but even if not, its dyadic form 
makes the decomposition a simple linear algebra problem. For example, 
suppose that at late times we have,
\begin{equation}
{\rm Tr}[S^n] \longrightarrow \left( \alpha {\rm Tr}[S]\right)^n \; ,
\end{equation}
for some positive constant $\alpha < 1$. The putative decomposition
would be,
\begin{equation}
S_{\vec{m} \vec{n}} = u_{\vec{m}} u_{\vec{n}} + {\Delta S}_{\vec{m}
\vec{n}} \; ,
\end{equation}
where the various traces obey,
\begin{equation}
u \cdot u \longrightarrow \alpha {\rm Tr}[S] \quad , \quad {\rm Tr}[
{\Delta S}] \longrightarrow (1-\alpha) {\rm Tr}[S] \quad , \quad
{\rm Tr}[{\Delta S}^n] \ll \left({\rm Tr}[S]\right)^n \; .
\end{equation}
Simply pick any mode function $v_{\vec{n}}$ with nonzero overlap --- 
it need not be close to $u_{\vec{n}}$ --- then contract into $S^2$
and divide by the trace of $S$,
\begin{equation}
{(S^2 v)_{\vec{n}} \over {\rm Tr}[S]} = \alpha u \cdot v \; u_{\vec{n}} 
+ {u^t {\Delta S} v \over {\rm Tr}[S]} \; u_{\vec{n}} + 
{u \cdot v \over {\rm Tr}[S]} ({\Delta S} u)_{\vec{n}} + 
{({\Delta S}^2 v)_{\vec{n}} \over {\rm Tr}[S]} \; .
\end{equation}
Only the first term can matter at late times so we recover 
$u_{\vec{n}}$ by normalizing and then multiplying by the square 
root of $\alpha {\rm Tr}[S]$.

\section{Summary and discussion}

We have calculated the gravitational back reaction on scalar-driven 
inflation using an invariant observable, to quadratic order in the
initial creation and annihilation operators and to leading
order in the long wavelength and slow roll approximations. No effect 
was found, contrary to previous work by ourselves and others which 
indicated a secular slowing of the expansion rate at this order 
\cite{MAB,ABM,AbWo1}. It is significant that the inclusion of 
stochastic effects played no role in this change. Under our 
approximations, secular back-reaction would enter through $\Phi(x)$,
the accumulated Newtonian potential from modes which have redshifted
into the infrared regime. However, all terms of order $\Phi$ and
$\Phi^2$ drop out of the Heisenberg operator ${\cal A}_{\rm inv}$,
before one has to choose between the alternatives of expectation 
values or stochastic samples. 

Another improvement in our analysis which had nothing to do with 
changing the result was the use of a scalar to measure the expansion 
rate. The entire difference from previous work is in fact attributable 
to defining the coordinate system so as to make the quantum scalar
vanish on surfaces of simultaneity. When these coordinates are used
even the gauge fixed expectation value of the metric fails to show 
significant back-reaction at one loop.

The null result was anticipated by Unruh who noticed that scalar 
mode solutions become pure gauge in the long wavelength limit 
\cite{Unruh}. This does not preclude back-reaction but it does 
rule out dependence upon $\Phi$ which fails to vanish in the long
wavelength limit. For example, spatial and certain temporal derivatives 
of $\Phi(x)$ can contribute. In fact there are such contributions
but they are negligible at quadratic order.

Although we have obtained a discouraging result about the possibility 
of a simple one-loop effect, our analysis does not invalidate the 
idea of gravitational back reaction on inflation. It would be fairer 
to say that what we have learned constrains the form any such effect 
can take. In particular one should not take the long wavelength 
limit. It is highly significant, in this regard, that the purely 
gravitational effect claimed at two loops \cite{TW1} does not
involve the long wavelength approximation in any way because the
locally de Sitter background is simple enough that the full propagator
can be worked out. In fact the gravitational response to the 
inflationary production of gravitons comes entirely from the
graviton kinetic energy and would vanish in the long wavelength
limit. 

The long wavelength approximation was also avoided in the effect
claimed at three loops for massless, minimally coupled $\varphi^4$
theory in a locally de Sitter background \cite{TW2}. In a subsequent
paper \cite{Obs3} we resolve the issue of coincident propagators in 
this model by a procedure of covariant normal ordering. The 
resulting theory exhibits a secular back-reaction which slows 
inflation in a manner that is unaltered by either the use of an 
invariant operator to measure expansion or by the inclusion of 
stochastic effects.

Although there was no need to consider stochastic effects in the
present work, Section 6 describes a formalism for including them 
in higher order processes which may show secular back-reaction. 
This formulation differs from the standard one \cite{stoch} in three 
ways. First, the focus is perturbative and local whereas previous 
previous treatments have been concerned with nonperturbative effects 
on the global geometry. A second difference is that past treatments 
incorporated stochastic degrees of freedom only as they experienced 
superadiabatic amplification. Although this is doubtless an excellent 
approximation for the sorts of global and nonperturbative issues that 
were being studied, we cannot afford to ignore the non-conservation 
of stress-energy implicit in continually injecting new degrees of 
freedom into the system. Therefore, the creation and annihilation 
operators for any mode we wish to treat stochastically are considered 
to be nonzero random numbers even on the initial value surface. The 
final difference is that we enforce all the perturbative field 
equations of the Einstein-scalar system so that the various modes are 
in gravitational interaction even on the initial value surface. 

An amusing consequence of all this is that stochastic effects provide
a nonperturbative proof that back-reaction must eventually become 
significant if nothing else stops inflation first. For it will be noted 
that, in our version, including stochastic effects at some time $\eta$ 
corresponds to a universe that began inflation at $\eta_0$ with a random 
collection of modes excited up to co-moving wave number $k = H(\eta) 
a(\eta)$. The hypothesis that inflation {\it never} slows amounts to 
the assumption that inflation can begin with the inital state populated 
to arbitrarily high wave number. Of course this is nonsense. Even if one
subtracts off the spatially averaged energy density --- as should probably 
be done --- what must actually happen for arbitrarily high excitations
is that random inhomogeneities produce a gravitational collapse.

Finally, we comment on the degree to which a random stochastic sample
should be expected to differ from its mean. Because secular back-reaction 
must manifest itself through the coherent superposition of an enormously 
large number of independent random variables, one can sometimes employ the 
methods of statistical mechanics. The exception is when back-reaction
involves an ordinary function of the local stochastic fields, for example
$\Phi^2(x)$. This can happen if the terrific inflationary expansion 
causes the average over the past lightcone to be weighted so that only
interactions just before the observation contribute effectively. In
that case stochastic fluctuation is not negligible, but the sign of the
effect is definite. In the other, ``nonlocal'' case, stochastic fluctuation
is negligible compared with the mean effect and one may as well use 
expectation values. Our suspicion is that improving upon the long 
wavelength approximation and improving on the expansion observable will 
result in secular back-reaction of the nonlocal sort. However, if there
should be local mixing, it is straightforward to untangle.

\vskip 1cm
\centerline{\bf Acknowledgments}

It is a pleasure to acknowledge stimulating and informative conversations 
with R. Bond, R. Brandenberger, T. Buchert, A. Guth, L. Koffman, A. Linde, 
D. Lyth, V. Mukhanov, E. Stewart and W. Unruh. We are also grateful to the 
University of Crete and to the Aspen Center for Physics for their hospitality 
during portions of this project. This work was partially supported 
by the Sonderforschungsbereich 375-95 f\"ur Astro-Teilchenphysik der
Deutschen Forschungsgemeinschaft, by DOE contract DE-FG02-97ER\-41029, 
by NSF grant 94092715 and by the Institute for Fundamental Theory.

\vskip 0.5cm
\noindent \appendix{{\bf \Large Appendix}}  
\vskip 0.5cm

In this Appendix we derive the infrared limits of the generalized 
Green's functions of Section 3.4. First notice that from Eq. (\ref{def:DI}) 
and the background identities we can derive the following useful 
``commutation'' relations:
\begin{eqnarray}
\partial_0 \widehat{\cal{D}}_B - \widehat{\cal{D}}_B \partial_0
& = & \frac{\kp^2}2 \vp_0' \vp_0'' \; , \label{a:cr1} \\
\vp_0' \widehat{\cal{D}}_B - \widehat{\cal{D}}_C \vp_0' 
= 2 \vp_0' \partial_0 \frac{\vp_0''}{\vp_0'} & , &
{\vp_0'} \widehat{\cal{D}}_C - \widehat{\cal{D}}_B {\vp_0'} 
= 2 \frac{\vp_0''}{\vp_0'} \partial_0 \vp_0' \; . \label{a:cr2} \\
\frac{1}{\vp_0'} \widehat{\cal{D}}_B - \widehat{\cal{D}}_C \frac{1}{\vp_0'} 
= - 2 \frac{\vp_0''}{{\vp_0'}^2} \partial_0 & , & \frac{1}{\vp_0'} 
\widehat{\cal{D}}_C - \widehat{\cal{D}}_B \frac{1}{\vp_0'} = - 2 
\partial_0 \frac{\vp_0''}{{\vp_0'}^2} \; . \label{a:cr3}
\end{eqnarray}
Using Eqs. (\ref{def:DI}) and (\ref{thetaI}) for 
$\widehat{\cal{D}}_B$ and $\widehat{\cal{D}}_C$ we have also the relations
\begin{eqnarray}
\label{a:cr6}
\widehat{\cal{D}}_B \frac{1}{{\vp_0'}^2} \partial_0 \vp_0' &=& 
\partial_0 \frac{1}{\vp_0'} \widehat{\cal{D}}_C - 
2 \frac{\vp_0''}{{\vp_0'}^2} \nabla^2 \; ,
\\ \label{a:cr7}
\widehat{\cal{D}}_C \frac{1}{\vp_0'} \partial_0 &=& 
\vp_0' \partial_0 \frac{1}{{\vp_0'}^2} \widehat{\cal{D}}_B + 
2 \frac{\vp_0''}{{\vp_0'}^2} \nabla^2 \; .
\end{eqnarray}
Dropping spatial derivatives and inverting implies the following
``inverse commutation'' relations,
\be
\frac{1}{\vp_0'} \partial_0 \vp_0' {\cal{D}}_C^{-1} \vp_0' = 
\vp_0' {\cal{D}}_B^{-1} \partial_0 \quad , \quad
\frac{1}{\vp_0'} \partial_0 {\cal{D}}_B^{-1} \vp_0' = 
{\cal{D}}_C^{-1} \vp_0' \partial_0 \frac{1}{\vp_0'} \; , \label{a:cr9}
\ee

Substituting (\ref{a:cr9}) into the right-hand-side of Eq. (\ref{eq:zzD}) 
we see that the $1/\nabla^2$ term cancels because,
\begin{eqnarray}
\lefteqn{\partial_0 {\cal{D}}_B^{-1} \partial_0 - \frac{\kp^2}4 \vp_0'
{\cal{D}}_C^{-1} \vp_0' = \vp_0' {\cal{D}}_C^{-1} \vp_0' \partial_0 
\frac{1}{{\vp_0'}^2} \partial_0 - \frac{\kp^2}4 \vp_0' {\cal{D}}_C^{-1} 
\vp_0' \; ,} \\
& & = \vp_0' {\cal{D}}_C^{-1} \left[ - {\cal{D}}_C + \frac{\kp^2}4 
{\vp_0'}^2 \right] \frac{1}{\vp_0'} - \frac{\kp^2}4 \vp_0'
{\cal{D}}_C^{-1} \vp_0' = -1 \; .
\end{eqnarray}
Similar manipulations reveal that the $1/\nabla^2$ terms cancel as
well in the expansions (\ref{eq:ffD})-(\ref{eq:zfD}).

The next order terms can be found by expanding $\widehat{\cal D}_I^{-1}$
in powers of the Laplacian. Doing this in expression (\ref{eq:zzD}) for 
$G_{zz}(x;x')$ yields,
\begin{equation}
G_{zz}(x;x') = \left[ - \partial_0 {\cal{D}}_B^{-1} {\cal{D}}_B^{-1} 
\partial_0 + \frac{\kp^2}{4} \vp_0' {\cal{D}}_C^{-1} {\cal{D}}_C^{-1} \vp_0' 
+ O(\nabla^2) \right] \delta^4 (x-x') \; .
\end{equation}
The term of order $\nabla^0$ can be simplified with the commutation 
relations (\ref{a:cr1}-\ref{a:cr3}),
\begin{eqnarray}
\lefteqn{- \partial_0 {\cal{D}}_B^{-1} {\cal{D}}_B^{-1} \partial_0 + 
\frac{\kp^2}{4} \vp_0' {\cal{D}}_C^{-1} {\cal{D}}_C^{-1} \vp_0' }
\nonumber \\
& & = - \vp_0' {\cal{D}}_C^{-1} \vp_0' \partial_0 \frac{1}{{\vp_0'}^2}
{\cal{D}}_B^{-1} \partial_0 + \frac{\kp^2}{4} \vp_0' {\cal{D}}_C^{-1} 
\frac{1}{\vp_0'} \frac{1}{\partial_0} {\vp_0'}^2 {\cal{D}}_B^{-1} 
\partial_0 \; , \\
& & = - \vp_0' {\cal{D}}_C^{-1} \vp_0' \partial_0 \frac{1}{{\vp_0'}^2}
{\cal{D}}_B^{-1} \partial_0 + \vp_0' {\cal{D}}_C^{-1} \frac{1}{\vp_0'} 
\frac{1}{\partial_0} ({\cal{D}}_B + \partial_0^2) {\cal{D}}_B^{-1} 
\partial_0 \; , \\
& & = - \vp_0' {\cal{D}}_C^{-1} \left( \frac{1}{\vp_0'} \partial_0 - 
2 \frac{\vp_0''}{{\vp_0'}^2} \right) {\cal{D}}_B^{-1} \partial_0 \nonumber \\
& & \qquad \qquad + \vp_0' {\cal{D}}_C^{-1} \frac{1}{\vp_0'} + \vp_0'
{\cal{D}}_C^{-1} \frac{1}{\vp_0'} \partial_0 {\cal{D}}_B^{-1} 
\partial_0 \; , \\
& & = \vp_0' {\cal{D}}_C^{-1} \frac{1}{\vp_0'} + 2 \vp_0' {\cal{D}}_C^{-1} 
\frac{\vp_0''}{{\vp_0'}^2} {\cal{D}}_B^{-1} \partial_0 \; ,
\end{eqnarray}
The last line, acting of a delta function, gives the long wavelength
limit of the $zz$ retarded propagator, Eq. (\ref{eq:zzir}). Similar reductions
pertain for the other Green's functions of the $(f,z)$ sector,
(\ref{eq:ffir})-(\ref{eq:zfir}).

\end{document}